\documentclass[twocolumn, resetfootnote]{aastex701}

\newcommand{\msun}{M$_{\sun}$}

\newcommand{\rstar}{R$_{\star}$}

\newcommand{\mstar}{M$_{\star}$}

\newcommand{\lacc}{L$_{\rm acc}$}

\newcommand{\msunyr}{\msun\,yr$^{-1}$}

\newcommand{\teff}{$T_{\rm eff}$}
\newcommand{\tshock}{$T_{\rm shock}$}
\newcommand{\tmax}{$T_{\rm max}$}

\newcommand{\halpha}{H$\alpha$}
\newcommand{\hbeta}{H$\beta$}
\newcommand{\brgamma}{Br$\gamma$}
\newcommand{\mdot}{$\dot{M}$}
\newcommand{\ri}{$R_{\rm i}$}
\newcommand{\rw}{$W_{\rm r}$}

\newcommand{\paperI}{\cite{Pittman2025}}

\newcommand{\rco}{$R_{\rm co}$}

\newcommand{\rd}{$R_{\rm dust}$}

\newcommand{\Beq}{$B_{\rm \star,eq}$}
\newcommand{\Bpolar}{$B_{\rm \star,polar}$}
\newcommand{\imag}{$i_{\rm mag}$}

\newcommand{\omegas}{$\omega_s$}
\newcommand{\omegasMed}{$\omega_{s{\rm ,med}}$}
\newcommand{\omegasMean}{$\omega_{s{\rm ,mean}}$}

\newcommand{\prot}{$P_{\rm rot}$}
\newcommand{\porb}{$P_{\rm orb}$}
\newcommand{\vrot}{$v_{\rm rot}$}
\newcommand{\vBr}{$v_{\rm breakup}$}

\newcommand{\degrees}{$^\circ$}

\newcommand{\dt}{$\Delta t$}

\newcommand{\MAD}{$\sigma_{\rm MAD}$}
\newcommand\meanOmega{0.51}
\newcommand\medianOmega{0.34}
\newcommand\MADOmega{0.19}

\newcommand\medianOmegaWind{0.45}

\newcommand\medianOmegaNoWind{0.22}
\newcommand\meanProt{5.2}
\newcommand\medianProt{4.9}
\newcommand\stdevProt{2.6}
\newcommand\MADProt{1.5}

\newcommand\be {\begin{equation}}
\newcommand\en{\end{equation}}

\def\fp1{$f_{p1}$}

\usepackage{tikz}
\usepackage[percent]{overpic}
\usepackage{subcaption}

\usepackage[utf8]{inputenc}
\usepackage{graphicx}
\usepackage{amsmath,amssymb}
\usepackage{breqn}
\usepackage{booktabs}
\usepackage{longtable}
\usepackage{lipsum}
\usepackage{amsmath}
\usepackage{float}
\usepackage{comment}

\usepackage{colortbl}
\usepackage{color}

\definecolor{propeller_color}{HTML}{2C4900}
\newcommand{\propeller}[1]{{\color{propeller_color} #1}}
\definecolor{stable_color}{HTML}{859899}
\newcommand{\stable}[1]{{\color{stable_color} #1}}
\definecolor{unstablechaotic_color}{HTML}{26A1A6}
\newcommand{\unstablechaotic}[1]{{\color{unstablechaotic_color} #1}}
\definecolor{unstableordered_color}{HTML}{94CC89}
\newcommand{\unstableordered}[1]{{\color{unstableordered_color} #1}}

\shorttitle{ODYSSEUS Star-Disk Connection}
\shortauthors{Pittman et al.}

\graphicspath{{./}{figures/}}

\begin{document}

\title{The ODYSSEUS Survey. Using accretion and stellar rotation to reveal the star-disk connection in T Tauri stars}

\correspondingauthor{Caeley V. Pittman}
\email{cpittman@bu.edu}

\author[0000-0001-9301-6252]{Caeley V. Pittman}\altaffiliation{NSF Graduate Research Fellow}
\affiliation{Department of Astronomy, Boston University, 725 Commonwealth Avenue, Boston, MA 02215, USA}
\affiliation{Institute for Astrophysical Research, Boston University, 725 Commonwealth Avenue, Boston, MA 02215, USA}
\email{cpittman@bu.edu}

\author[0000-0001-9227-5949]{Catherine C. Espaillat}
\affiliation{Department of Astronomy, Boston University, 725 Commonwealth Avenue, Boston, MA 02215, USA}
\affiliation{Institute for Astrophysical Research, Boston University, 725 Commonwealth Avenue, Boston, MA 02215, USA}
\email{cce@bu.edu}

\author[0000-0003-3616-6822]{Zhaohuan Zhu}
\affil{Department of Physics and Astronomy, University of Nevada, Las Vegas, 4505 S. Maryland Pkwy, Las Vegas, NV 89154, USA} 
\affil{Nevada Center for Astrophysics, University of Nevada, Las Vegas, 4505 S. Maryland Pkwy., Las Vegas, NV 89154-4002, USA} 
\email{zhaohuan.zhu@unlv.edu}

\author[0000-0003-4507-1710]{Thanawuth Thanathibodee}
\affiliation{Department of Physics, Faculty of Science, Chulalongkorn University, 254 Phayathai Road, Pathumwan, Bangkok 10330, Thailand}
\email{Thanawuth.T@chula.ac.th}

\author[0000-0003-1639-510X]{Connor E. Robinson}
\affiliation{Division of Physics and Astronomy, Alfred University, 1 Saxon Drive, Alfred, NY 14802, USA}
\email{robinsonc@alfred.edu}

\author[0000-0002-3950-5386]{Nuria Calvet}
\affiliation{Department of Astronomy, University of Michigan, 311 West Hall, 1085 S. University Avenue, Ann Arbor, MI 48109, USA}
\email{ncalvet@umich.edu}

\author[0000-0001-7157-6275]{\'Agnes K\'osp\'al}
\affiliation{Konkoly Observatory, HUN-REN Research Centre for Astronomy and Earth Sciences, MTA Centre of Excellence, Konkoly-Thege Mikl\'os \'ut 15-17, 1121 Budapest, Hungary}
\affiliation{Institute of Physics and Astronomy, ELTE E\"otv\"os Lor\'and University, P\'azm\'any P\'eter s\'et\'any 1/A, 1117 Budapest, Hungary}
\affiliation{Max Planck Institute for Astronomy, K\"onigstuhl 17, 69117 Heidelberg, Germany}
\email{kospal.agnes@csfk.org}

\submitjournal{ApJ}
\received{August 7, 2025}
\revised{September 2, 2025}
\accepted{September 3, 2025}

\begin{abstract}

Classical T Tauri stars (CTTS) exhibit strong variability over timescales of minutes to decades. However, much theoretical work assumes that CTTS are in stable spin states.
Here, we test expectations for CTTS angular momentum regulation by comparing star and disk rotation. We measure stellar rotation periods and disk corotation radii (\rco) for 47 CTTS from the HST ULLYSES sample. We compare \rco\ to the magnetospheric truncation radii (\ri) and show that most CTTS are in the spin-up regime based on model predictions, which may indicate efficient angular momentum loss processes. We find evidence of magnetospheric outflows and episodic accretion, and our observations are consistent with the presence of accretion-powered stellar winds.
We confirm predictions that \ri\ is variable over timescales of days, causing some CTTS to cross accretion stability regime boundaries.
We characterize light curve morphologies and confirm that our inclined CTTS with \ri$\sim$\rco\ show dipper light curves, consistent with expectations from disk warp models.
However, dippers occur at all values of \ri/\rco, suggesting that they do not need to be near the propeller regime.
Finally, we show that our measured \ri\ locations are consistent with observed ultra-short-period planet (USP) semi-major axes. If USPs are stable against tidal dissipation, as has been suggested in the literature, then our work provides a plausible USP formation channel.
These results show that the star-disk connection produces a large variety of accretion and stellar spin configurations, most of which are likely not in equilibrium.

\end{abstract}


\section{Introduction} \label{sec:intro}

Classical T Tauri stars (CTTS) are low-mass ($\rm M_\star<2$~\msun) pre-main-sequence stars that are actively accreting from their surrounding protoplanetary disks. While accretion and gravitational contraction can act as spin-up processes,
CTTS are known to maintain relatively slow rotation rates throughout their 5--10~Myr lifetimes \citep{Irwin2009,Bouvier2014}.
This implies the existence of efficient angular momentum removal mechanisms that balance the spin-up torques.
Early work suggested that the inner disk might be ``locked'' to the star, providing a spin-down mechanism for the CTTS \citep{GhoshLamb1979a,GhoshLamb1979}, but later work has shown that disk locking is inefficient for regulating CTTS rotation \citep[e.g.][and references therein]{AgapitouPapaloizou2000,Uzdensky2002,MattPudritz2005,Herbst2007,ZanniFerreira2009}.
More probable sources of angular momentum regulation include magnetospheric outflows \citep{Shu2000,romanova2009,ZanniFerreira2013,romanova2018,Takasao2022,Ireland2022,Gaidos2024,Zhu2025}, episodic accretion and outflow cycles \citep{Audard2014}, and accretion-powered stellar winds \citep[APSWs;][]{MattPudritz2005,MattPudritz2008a,MattPudritz2008b}.

There are two critical radii in the inner disk of CTTS that are predicted to determine whether a system is in the spin-up, spin-down, or equilibrium spin state: the magnetospheric truncation radius \ri\ and the corotation radius \rco.
\ri\ is the radius at which the magnetic pressure from the magnetosphere balances the ram pressure from the accretion disk. It marks the discontinuity within which most disk matter will accrete onto the star along magnetic field lines or become ejected in an outflow.
\rco\ marks the location at which the disk's Keplerian angular velocity matches that of the star, such that faster-spinning stars will have smaller \rco/\rstar\ values.
The corotation radius can be calculated from the stellar mass (\mstar) and rotation period (\prot) according to ${R_{\rm co}=({\rm GM}_\star P_{\rm rot}^2}/4\pi^2)^{1/3}$.

The relationship between \ri\ and \rco\ has been parameterized by the ``fastness parameter,'' \omegas$=(R_{\rm i}/R_{\rm co})^{3/2}$ \citep[][]{Ghosh2007}, which is equivalent to the ratio of the stellar angular velocity to the Keplerian angular velocity at \ri\ (\omegas$=\Omega_\star/\Omega_{\rm K,R_i}$). 
Observationally, \ri\ can be measured by fitting an accretion flow model to velocity-resolved \halpha\ emission lines \citep[e.g.,][]{muzerolle01,thanathibodee23,wendeborn24c,Pittman2025}, and \rco\ can be measured by obtaining \prot\ from CTTS light curves.

Three-dimensional magnetohydrodynamic (3D MHD) simulations have found that \omegas\ is the most important determinant of the accretion stability and angular momentum evolution of a given system \citep[e.g.,][]{blinova16,romanova2018,Zhu2025}. 
The propeller regime, in which the star spins faster than the Keplerian disk at \ri, is defined by $\omega_s>1$; the stable regime, in which accretion occurs through long-lived accretion flow configurations, is expected to operate where  $1\gtrsim\omega_s\gtrsim0.6$; the unstable chaotic regime, in which accretion occurs through stochastic Rayleigh-Taylor protrusions into the magnetosphere, is predicted where $0.6\gtrsim\omega_s\gtrsim0.45$; and, finally, the unstable ordered regime, in which accretion occurs through one or two ordered accretion ``tongues,'' is expected where $\omega_s\lesssim0.45$ \citep{blinova16,Zhu2025}.

The fastness parameter should also dictate the latitude at which the accreting material reaches the star.
In the propeller regime, accretion is expected to be weak and occur along high-latitude open magnetic field lines that do not act as a centrifugal barrier \citep{Romanova2004b,romanova2018,Takasao2022,Zhu2025}. In the stable regime, accretion occurs at intermediate latitudes, and in the unstable regime, accretion is expected to occur closer to the stellar equator \citep{blinova16,Zhu2025}.

If a CTTS's magnetic moment is aligned with its rotation axis within $\sim$5\degrees\ and the disk is stable, the hotspot is expected to be a ring surrounding the pole \citep{KulkarniRomanova2013}. However, at more typical magnetic obliquities of 10\degrees--20\degrees\ \citep{Donati1997,Donati2010,McGinnis2020,Nelissen2023a,Nelissen2023b,donati2024}, the hotspot is expected to be an arc with an azimuthally-stratified density gradient \citep{KulkarniRomanova2013}.
Even in the case of low magnetic obliquity, the hotspot may exhibit complex asymmetry due to MHD instabilities that cause non-axisymmetric mass loading onto the stellar magnetic field lines \citep[][]{zhu2024,Zhu2025}.

As highly magnitized low-mass stars, CTTS are also known to have cold magnetic starspots.
The combination of accretion hotspots and cold starspots results in periodic, rotation-induced light curve variability as the spot emission is modulated at the stellar rotation period.
This permits the measurement of \prot\ given a sufficient temporal baseline and cadence of observations \citep[e.g.,][]{Bouvier1986,Bouvier1993,Batalha1998,percy10,Cody2010,Mellon2017,Rebull2018,serna21,Cody2022}. 
The month-long TESS sectors, which are taken at high temporal resolution, are sufficient to constrain CTTS rotation periods of $\sim$1--10~days \citep[][]{Bouvier2014,Smith2023,Boyle2025} in systems with typical accretion rates (i.e., systems in which strong and stochastic accretion emission does not dominate the light curve).

Magnetospheric accretion is observed to be variable on timescales of minutes to years, both from intrinsic variability and rotational modulation of the visible emission \citep[for a review, see][]{fischer23_ppvii}. 
The specific morphology of light curves can indicate the stability of the accretion system \citep{blinova16,Sousa2016}, the presence of intervening disk material such as disk warps or dusty magnetospheres \citep{Bouvier1999,McGinnis2015,Nagel2025}, and intrinsic system parameters such as magnetic field geometry, viewing inclination, and turbulence \citep{Cody14,Cody18,Robinson2021}.

In \cite{Pittman2025}, we performed self-consistent modeling of the accretion flow and shock in the HST ULLYSES sample.
We found that magnetospheric truncation radii are typically much smaller than has been previously assumed, with a median value of 2.8~\rstar.
In this work, we examine the connection between accretion, stellar rotation, outflows, and variability morphologies by combining the results of \cite{Pittman2025} with new analysis of TESS observations of the same CTTS. 
In Section~\ref{sec:analysis}, we describe the sample, data, analysis methods, and results. 
In Section~\ref{sec:discussion}, we place our results in context with expectations for angular momentum evolution in CTTS, and then discuss implications for short-period exoplanet formation. We conclude with a summary in Section~\ref{sec:summary}.  

\section{Analysis and Results} \label{sec:analysis}

The initial sample we use consists of the 66 CTTS with \ri\ measurements in \paperI. These \ri\ values come from modeling \halpha\ emission line profiles with the accretion flow model of \cite{hartmann94} and \cite{muzerolle98a,muzerolle01}. In this work, we measure \rco\ to calculate the fastness parameter \omegas\ and classify the systems into the expected accretion stability regimes of \cite{blinova16}. To measure \rco, we perform periodogram analysis of TESS light curves to obtain \prot.  
All 66 CTTS in the initial sample have at least one sector of TESS observations, but issues of crowding and stellar multiplicity reduce the number of usable targets to 53 CTTS.
We retain 7 multiple systems whose primary components dominate the TESS light curves (CS~Cha~A, CVSO~109~A, DK~Tau~A, Sz~19~A, Sz~68~A, and UX~Tau~A) or are spatially resolved from companions (DG~Tau~A). We also retain 2MASSJ11432669-7804454, which has a newly-discovered visual companion at 0\farcs165,\footnote{The system is unresolved in both HST and VLT observations; see \url{https://ullyses.stsci.edu/ullyses-dr3.html} for details.} because its light curve shows no evidence of a secondary periodic source.

The 53 CTTS with usable TESS light curves are given in Table~\ref{tab:results}, along with their stellar masses (\mstar), magnetospheric viewing inclinations (\imag), and accretion rates (\mdot) from \paperI. Table~\ref{tab:results} also presents the QM variability metrics, \prot, \rco, \ri, and \omegas\ from this work for each CTTS.
Figure~\ref{fig:RadiusFigure_twocol0} in Appendix~\ref{Appsec:app_flowresults} visualizes all 66 individual system configurations, including their stellar parameters, accretion flow configurations, viewing angles, and \rco\ locations when available.

\begin{figure*}[t]
    \centering
    \digitalasset
    \includegraphics[trim=7pt 7pt 7pt 6pt, clip, width=0.26\textwidth]{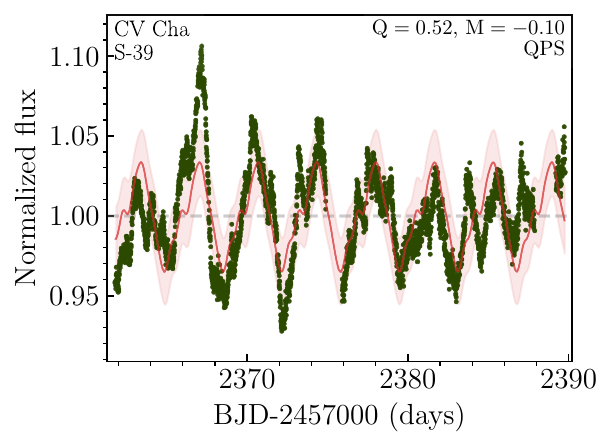}
    \includegraphics[trim=22pt 7pt 7pt 6pt, clip, width=0.24\textwidth]{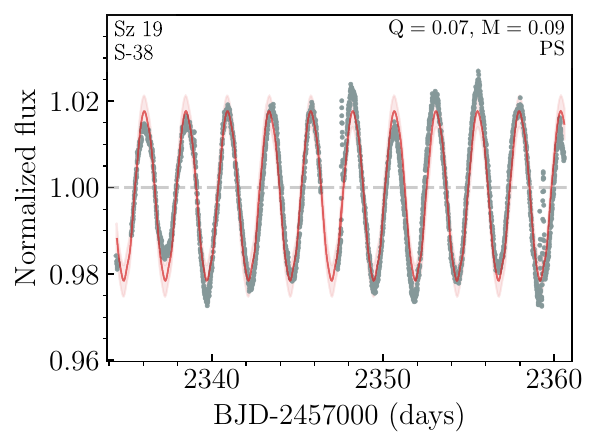}
    \includegraphics[trim=22pt 7pt 7pt 6pt, clip, width=0.24\textwidth]{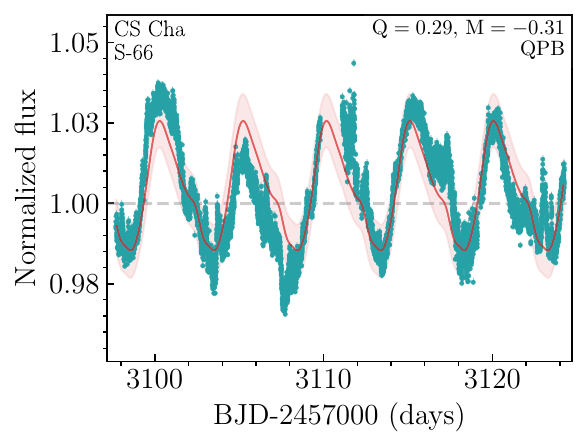}
    \includegraphics[trim=22pt 7pt 7pt 6pt, clip, width=0.24\textwidth]{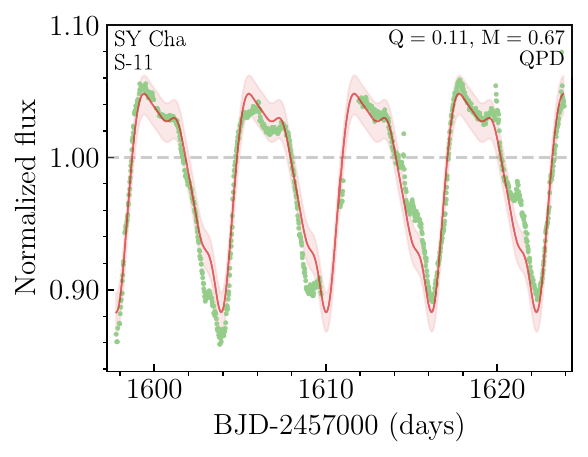}
    \caption{Example TESS light curves colored by predicted accretion stability regime. The associated QM metrics are in the upper right, and the TESS sector is in the upper left. A Gaussian process with a periodic kernel fixed to \prot\ from Table~\ref{tab:results} is shown in red, and its standard deviation is indicated by the shaded red region. All light curves are available in the figure set in the online materials.}
    \label{fig:TESS_short}
\end{figure*}

\subsection{Measuring $P_{\rm rot}$ and $R_{\rm co}$}\label{subsec:ProtRco}

We first use \texttt{TESSExtractor} \citep{brasseur19,serna21} to extract TESS image cutouts and perform aperture photometry for all $\sim$27~day TESS sectors of each target.
When the strong 13.7~day lunar synchronous signal appears in a light curve \citep[which is caused by scattered light from the Earth and Moon, see][]{Hattori2022}, we use the \texttt{TESSExtractor} de-trending tool to perform manual systematics correction.
TESS sectors are manually excluded if the CTTS falls at the edge of the detector, which produces unusable data.
We increase the signal-to-noise of each light curve by binning it into 2-hour bins, weighting each point according to its measurement uncertainty.
Then, we subtract a second-order polynomial from the light curves to remove secular brightness changes.

To measure \prot, we perform a 200-step bootstrapped Lomb-Scargle \citep[LS;][]{lomb76,scargle82} periodogram analysis using the \texttt{auto_period} function in \texttt{lightcurvy}, which was also used by \cite{wendeborn24b}.\footnote{\url{https://github.com/wendeborn8/lightcurvy}} In each step, \texttt{auto_period} generates a synthetic light curve drawn from a normal distribution with a mean given by the observed values and a standard deviation given by the photometric uncertainties. Then, we select 63\% of the points randomly and apply the LS algorithm. We obtain a final periodogram by choosing the 50th percentile of the 200 bootstrapped periodograms (see Figure~\ref{fig:LS_SF} in Appendix~\ref{Appsec:per_val}).

We look at all periods with power greater than 0.1 and ignore any that correspond to known TESS aliases.
When multiple sectors with good quality data are available, we choose the period that is present across multiple sectors to reduce the chance of tracing shorter-lived oscillations.
If the best period differs between sectors, periodograms are combined to determine a period that is consistent with both within their uncertainties.
When the best period is unclear from LS, we apply phase-dispersion minimization (PDM) to find the period that is consistent with both LS and PDM.
We ensure that our derived periods are not artifacts of the observational cadence by calculating the window function and confirming that its peaks do not produce the measured period.

For TESS periods that are abnormally short ($\sim$1~day) or long ($>$10~days), we further confirm the periods by using \texttt{lightcurvy} to find periods in all available photometry from the ASAS-SN, WISE, and Gaia archives. 
As a final validation step, we calculate the first-order structure function \citep[SF;][]{Simonetti1985} for each sector and visually confirm that the SF tends to be low at integer multiples of the period (e.g., $1P$, $2P$, $3P$) and high at half-integer multiples of the period (e.g., $P/2$, $3P/2$, $5P/2$; see Figure~\ref{fig:LS_SF} in Appendix~\ref{Appsec:per_val}). 
Figure~\ref{fig:TESS_short} shows four example TESS light curves and their best-fit Gaussian processes\footnote{As implemented in \texttt{QMy} using a periodic kernel from \texttt{sklearn.gaussian_process.kernels.ExpSineSquared()}. The \texttt{QMy} code is publicly available on Github at \url{https://github.com/wendeborn8/QMy}. This was also used by \cite{wendeborn24b}.} using the measured periods. Light curves for the complete sample are included in Figure Set 1 in the online materials.

We estimate period uncertainties using the strength and certainty of the period's LS periodogram peak in the following manner. First, we fit a skew-normal function \citep[which accounts for asymmetric uncertainties,][]{Azzalini2009SkewNorm} to the LS peak in each sector of observation. The amplitude ($A$) and variance ($\sigma$) of the peak indicate the strength and certainty of the period measurement, respectively.
In order to combine the measurements from individual sectors, we create a distribution of likely periods by taking random draws from each sector's skew-normal distributions. We take a total of 1000 draws, with the number of draws taken from each TESS sector proportional to $A$/$\sigma^2$ to give more weight to stronger, more certain period measurements. 
We obtain the lower and upper period uncertainties by taking the 16th and 84th percentiles of the 1000 period draws.

Next, we calculate \rco\ by assuming that the derived periods correspond to the stellar rotation period, using ${\rm R_{co}=(GM_\star P_{rot}^2}/4\pi^2)^{1/3}$.
Stellar masses come from \cite{Pittman2025}. They were determined using the methods described in \citet{manara21} and have an uncertainty of 0.1~dex. We estimate \rco\ uncertainties by propagating those of \mstar\ and \prot, and we find that the \mstar\ uncertainties typically dominate.

\subsubsection{Reliability of $P_{\rm rot}$, $R_{\rm co}$, and $\omega_s$ measurements} \label{subsec:ProtReliability}

We expect our TESS periods to be reliable, as \cite{Boyle2025} compared stellar rotation periods measured from TESS with those from K2 (with its longer baseline allowing detection of periods up to $\sim$100~days) and found that TESS periods have typical uncertainties below 3\% for \prot$<$10~days (which applies to 96\% of our sample and the majority of CTTS in general, e.g., \citealt{Bouvier2014,venuti17,Smith2023}).

To measure \rco, we assume that the measured period corresponds to the stellar rotation period.
\cite{blinova16} show that this assumption should be valid for systems with $\omega_s\gtrsim0.45$ (i.e., those in the propeller, stable, or unstable chaotic regimes). 
However, for small magnetic obliquities ($\sim$5\degrees), CTTS in the unstable ordered regime ($\omega_s\lesssim0.45$) are expected to accrete via one or two ordered tongues that rotate at the frequency of the inner disk \citep{blinova16}. 
This could impact the rotation periods we infer from light curves, as a periodic signal at the Keplerian inner disk frequency would be shorter than the true stellar rotation period in the unstable accretion regime \citep[see, e.g.,][]{Armeni2024}.

If $\rm P_{disk}<P_\star$, the unstable ordered systems would be pushed even further into the unstable ordered regime, as their true stellar rotation rates would be slower than our inferred rates (making \rco\ larger and \omegas\ smaller).
Still, the \cite{blinova16} Fourier analyses of unstable ordered systems for models with magnetic obliquities that are more typical for CTTS \citep[$\sim$$10^\circ-20^\circ$,][]{Donati1997,Donati2010,McGinnis2020,Nelissen2023a,Nelissen2023b,donati2024} indicate that the dominant frequency corresponds to $\rm P_\star$ rather than $\rm P_{disk}$. 
Therefore, $\rm P_{disk}$ effects should dominate only in low-\omegas\ systems with low magnetic obliquity. Even then, contributions from $\rm P_{disk}$ would not change the systems' classification into the unstable ordered regime because a smaller \omegas\ would still fall into $\omega_s\lesssim0.45$.

The \cite{blinova16} models do not include starspots when obtaining their periodograms. However, cool starspots are the primary source of rotation-induced periodicity in non-accreting T Tauri stars (weak-lined T Tauri stars; WTTS), and CTTS often show comparable amplitudes of variability as WTTS in the TESS wavelength range, which is centered at $I$-band \citep[][]{Hoffmeister1965,Herbst1994}. Except in the case of very high accretors, starspots will produce detectable periodic signals at $\rm P_\star$, and therefore contribute to reliable \rco\ measurements.

The final component needed to obtain a reliable \omegas\ measurement is a reliable \ri\ measurement. In \cite{Pittman2025}, we measured \ri\ using the accretion flow model of \cite{hartmann94} and \cite{muzerolle98a,muzerolle01}. This model assumes that accretion occurs through an axisymmetric flow that reaches the stellar surface in a ring around the magnetic moment. This geometry implies that the magnetic moment is aligned with the stellar rotation axis (i.e., zero magnetic obliquity) and is perpendicular to the plane of the inner gas disk. As discussed in Section~\ref{sec:intro}, 3D MHD simulations show that accretion is rarely axisymmetric, and observations show that magnetic obliquity is common. 
There are two primary concerns regarding the influence of non-axisymmetry in deriving \ri\ with an axisymmetric model: first, the derived \ri\ may depend on the rotation phase at which the system is viewed; second, it may depend on the specific morphology of the flows (i.e., one-stream, two-stream, or more chaotic configurations). 
Non-axisymmetric accretion could produce velocity shifts for which an axisymmetric model cannot account. If such shifts are significant, they should become apparent through rotational modulation.

\cite{Kurosawa2013} studied the rotational modulation of hydrogen emission lines in stable and unstable systems using 3D MHD models. 
Their model line profiles trace phases in which the star is mostly visible (with accretion flows in the plane of the sky or behind the star), as well as phases in which the star is mostly obscured (with flows in front of the star along the line of sight; see their Figures~3 and 4).
While rotational modulation is apparent for lines such as \hbeta\ and \brgamma\ in their models, especially in the redshifted absorption component, \halpha\ variability is minimal (see their Figure~A1).
\cite{Kurosawa2013} attribute this to the fact that \halpha\ emission comes from the full accretion flow, rather than only near the stellar surface, and its optical thickness produces Stark line broadening that fills in any redshifted absorption components \citep{muzerolle01}. Their models indicate that rotational modulation of non-axisymmetric flows has a small effect on the observed \halpha\ profile, which lends credence to our \ri\ measurements.

We can also compare the \halpha\ profiles from the stable and unstable configurations to examine the effects of the flow morphology. When viewed at a phase when the star is fully obscured, the unstable configuration approximates the axisymmetric flow because it has 4-5 accretion streams that are spread in azimuth (see phase 1.04 in \citealt{Kurosawa2013}).
Conversely, the stable regime has only one stream per hemisphere, which maximizes the effects of asymmetry.
\cite{Kurosawa2013} find that the two sets of profiles have essentially the same peak flux. Their primary difference lies in their line widths, with the unstable profile being $\sim$85~km/s broader than that of the stable profile.

\begin{figure}
    \centering
    \includegraphics[width=\linewidth]{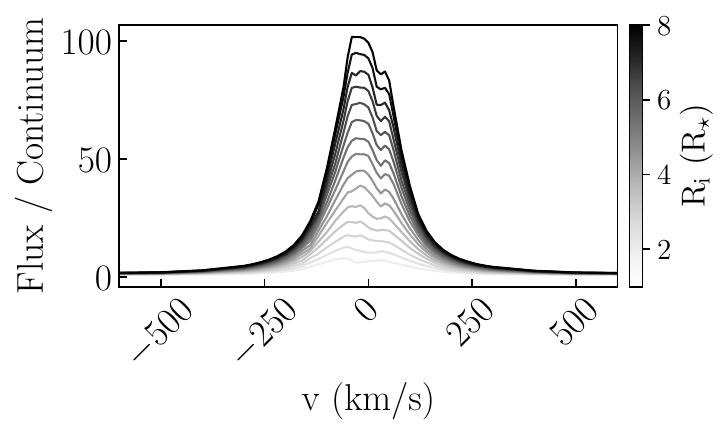}
    \caption{\halpha\ profile dependence on \ri, spanning from 2 to 8~\rstar. This is taken from the model grid in \cite{Pittman2025} for VZ~Cha.}
    \label{fig:Ri_dependence}
\end{figure}

To test whether this profile difference would influence our derived \ri, we plot axisymmetric model line profiles as a function of \ri\ in Figure~\ref{fig:Ri_dependence}. Here we use the model grid for VZ~Cha from \cite{Pittman2025}, fixing \rw, \mdot, \tmax, and \imag\ but varying \ri\ from 2.0 to 8.0~\rstar\ in steps of 0.4~\rstar.\footnote{We fix \rw\ to 0.9~\rstar, \mdot\ to $9.3\times10^{-9}$~\msunyr, \tmax\ to 8400~K, and \imag\ to 57\degrees. This matches the inclination of 60\degrees\ used in \cite{Kurosawa2013}.}
This shows that the primary effect of \ri\ is to increase the overall line flux, with a typical peak flux increase of 5 normalized flux units per \ri\ step. We can thus conclude that our \ri\ measurements should not be strongly affected by the specific morphology of the accretion streams, and our associated \omegas\ measurements should be reliable.

\subsection{Measuring Q \& M}\label{subsec:QM}

Once we have obtained \prot, we classify each TESS sector's morphology using the QM indices introduced in \cite{Cody14}.
The Q metric indicates whether a light curve is periodic, quasiperiodic, or stochastic, and the M metric indicates whether a light curve is dippy, bursty, or symmetric. We calculate these following \cite{Robinson2021} and \cite{wendeborn24b} using \texttt{QMy}.
In short, to measure Q, we first determine the smoothed periodic component by fitting a Gaussian process (GP) with a periodic kernel
to the phase-folded TESS data. We then subtract this periodic GP from the data to obtain the residual non-periodic components. The periodicity metric is calculated as ${\rm Q}=({\rm rms_{resid}^2} - \sigma^2) / ({\rm rms_{raw}^2} - \sigma^2)$, where $\rm rms_{resid}$ is the rms of the residuals, $\rm rms_{raw}$ is the rms of the original data, and $\sigma$ is the fractional uncertainty of the data. 

The specific values of Q that demarcate the periodic, quasi-periodic, and stochastic regimes depend on the source of the photometry, with Q being affected by the uncertainty estimation and observational cadence \citep{Cody14,Cody18,Bredall2020,Hillenbrand2022}. 
\cite{Cody14} found that CoRoT light curves are periodic for ${\rm Q}<0.11$, quasi-periodic up to ${\rm Q}=0.61$, and stochastic for higher Q. \cite{Cody18} found a higher upper boundary of ${\rm Q}=0.85$ for K2 light curves.
\cite{Hillenbrand2022} applied the analysis to ground-based ZTF light curves and needed to inflate photometric uncertainties to cause the Q distribution to extend to values higher than 0.65. Through visual inspection, they found Q boundaries of 0.45 and 0.87.

\cite{Hillenbrand2022} suggested that Q measurements should be scaled to span the [0--1] range, and the Q boundaries should be chosen using visual inspection. For this reason, we will discuss our chosen Q boundaries when we present our QM analysis results in Section~\ref{subsec:VariabilityMetrics}.
For each target, we confirm that at least one TESS sector is classified as periodic or quasi-periodic, rather than purely stochastic, when assuming our measured period.

To measure M, we perform 5$\sigma$ clipping on the data to remove outliers, and then calculate the mean of the top and bottom deciles of the light curve, $\langle d_{10\%}\rangle$. The symmetry metric is then calculated as ${\rm M}=-(\langle d_{10\%}\rangle-d_{\rm med})/\sigma_d$, where $d_{\rm med}$ and $\sigma_d$ are the median and rms of the sigma-clipped light curve. The negative sign applies to calculations in terms of flux, whereas the original metric was derived for observations in magnitudes. A light curve is symmetric for M between $\pm0.25$, a burster for ${\rm M}<-0.25$, and a dipper for ${\rm M}>0.25$, in accordance with the M boundaries found by \cite{Cody14}, \cite{Cody18}, and \cite{Hillenbrand2022}.

\begin{figure}[t]
    \centering
    \includegraphics[width=0.49\linewidth]{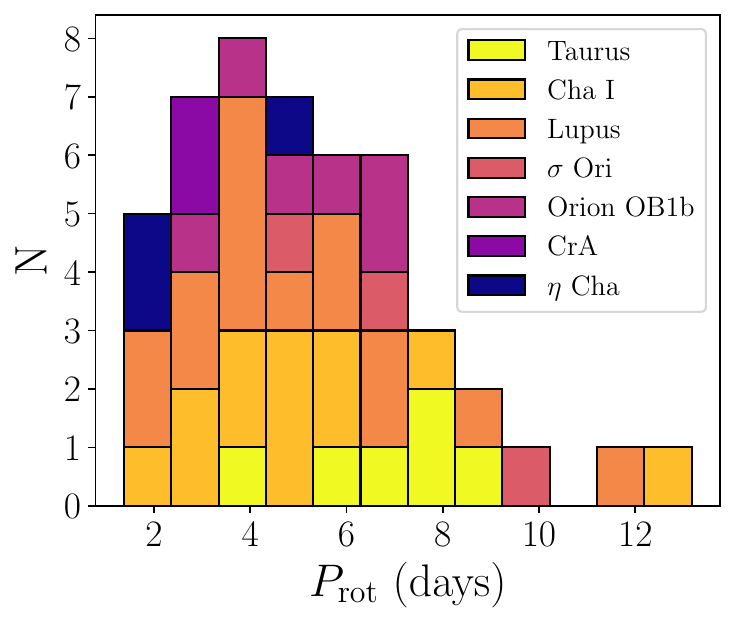}
    \includegraphics[width=0.49\linewidth]{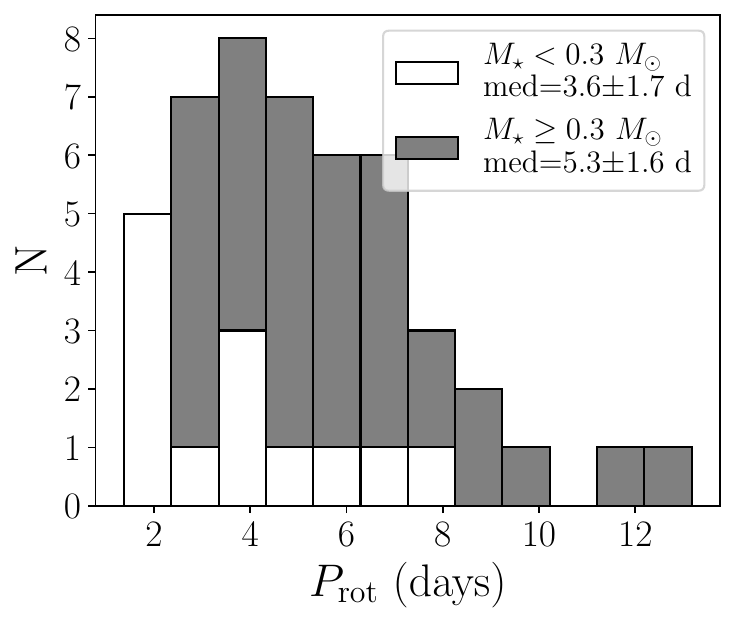}\\
    \includegraphics[width=0.49\linewidth]{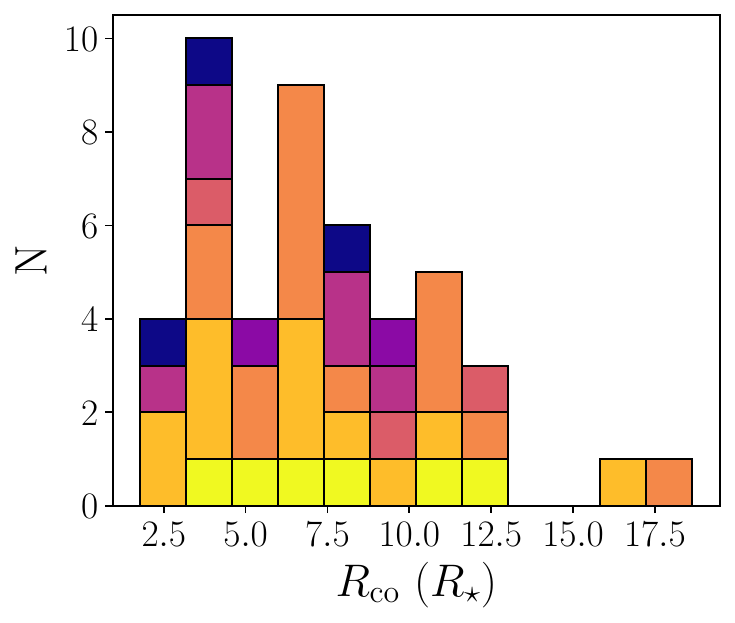}
    \includegraphics[width=0.49\linewidth]{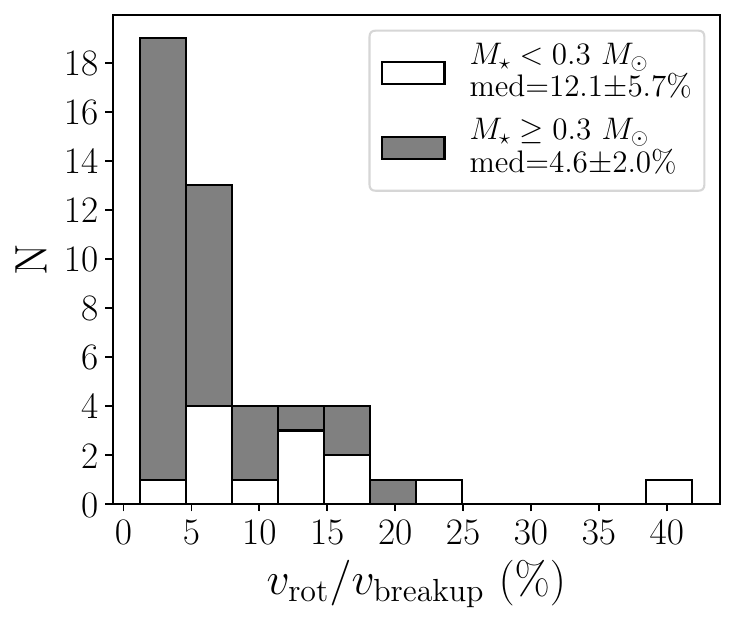}
    \caption{Histograms of stellar rotation results. Colors indicate the host region in order of approximate region age from youngest (yellow) to oldest (blue). White and gray indicate stellar mass bins. The top panels show stellar rotation periods measured from TESS light curves, and the bottom show the associated corotation radii and rotation velocities with respect to the stellar breakup velocities.
    }
    \label{fig:flow_hists}
\end{figure}

\subsection{$P_{\rm rot}$, $R_{\rm co}$, and $\omega_s$ results} \label{subsec:RotResults}

We obtain periods for 47 CTTS, with a median value of $\medianProt$~days, a median absolute deviation (\MAD) of $\MADProt$~days, a mean of \meanProt~days, and a standard deviation of $\stdevProt$~days (see Figure~\ref{fig:flow_hists}, top left). This is consistent with the \prot\ distribution mean of 6.1$\pm$1.3~days found for a sample of 200 CTTS by \cite{venuti17}.
The median ratio of the associated stellar rotation velocities (\vrot) to the stellar breakup velocities (\vBr) is 6\%, in agreement with the $<$10\% found in previous work \citep[e.g,][and references therein]{Bouvier1993,MattPudritz2005,Serna2024}.
Our \prot\ measurements correspond to corotation radii between 1.8-18.6~\rstar, with a median/\MAD\ \rco\ of $6.5\pm2.5$~\rstar\ (Figure~\ref{fig:flow_hists}, bottom left).


While the \prot\ sample is relatively small, it shows a potential mass dependence in agreement with previous work \citep{Herbst2001,Herbst2007,Smith2023}. We find that the fully convective CTTS ($M_\star\lesssim0.3$~\msun) have a median/\MAD\ \prot\ of $3.6\pm1.7$~days, whereas those with $M_\star\geq0.3$~\msun\ have a median/\MAD\ of $5.3\pm1.6$~days  (Figure~\ref{fig:flow_hists}, top right). Additionally, lower mass CTTS tend to rotate at a higher percentage of \vBr\ ($12.1\pm5.7$\%) than higher mass CTTS ($4.6\pm2.0$\%; see Figure~\ref{fig:flow_hists}, bottom right). This mass dependence supports previous assertions that internal stellar structure plays an important role in angular momentum evolution.

\begin{figure*}
    \centering
    \includegraphics[width=\textwidth]{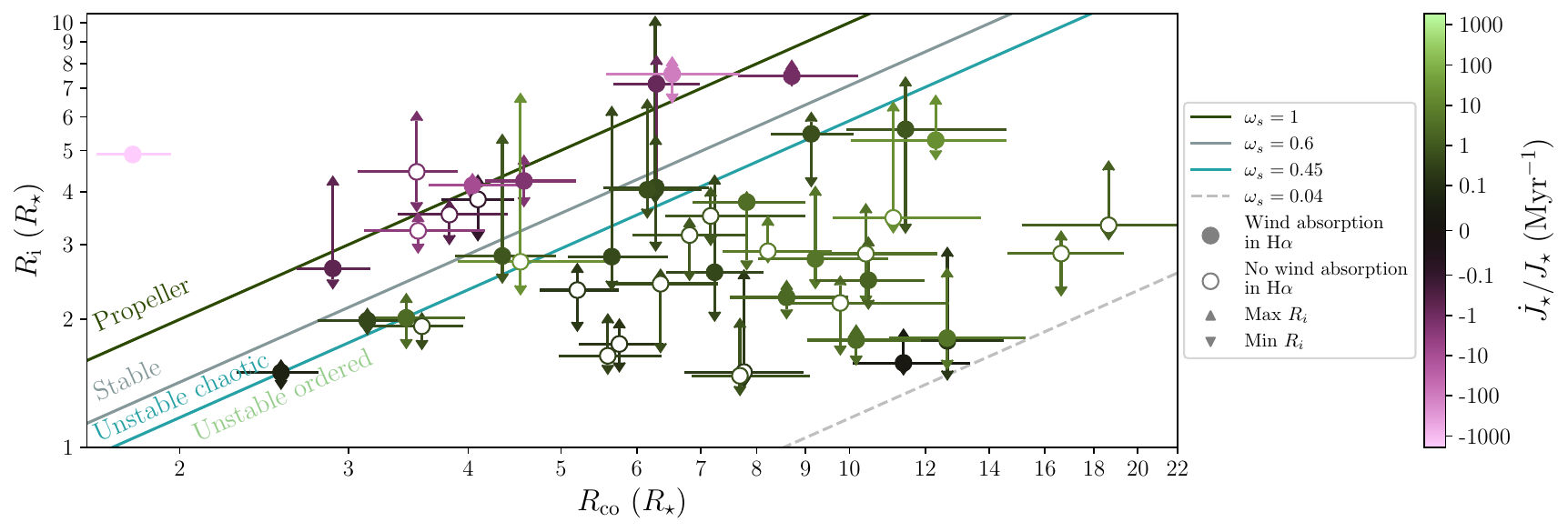}
    \caption{Inner magnetospheric truncation radii \ri\ versus corotation radii \rco\ for the 47 CTTS with successful \prot\ measurements. The scatter points give the \ri\ found from the weighted means of the individual epochs, the horizontal errorbars give the \rco\ uncertainty, and the vertical arrows span the space between the minimum and maximum \ri\ from the individual epochs. 
    The colorbar shows the fractional spin evolution rate, where bright green indicates the spin-up regime, bright pink indicates the spin-down regime, and dark colors indicate systems near spin equilibrium.
    Filled circles indicate objects that show blueshifted absorption in \halpha\ indicative of winds. Solid diagonal lines show the values of the fastness parameter \omegas\ at the cutoffs between accretion regimes (\omegas$=$1, 0.6, and 0.45) from \cite{blinova16} for a magnetic obliquity of 5\degrees. Note that the axes are logarithmic, which affects the apparent magnitude of the \rco\ uncertainty and \ri\ variability.
    }
    \label{fig:RiRco}
\end{figure*}

Figure~\ref{fig:RiRco} shows \ri\ versus \rco\ for our sample and demarcates the different stability regimes predicted by \cite{blinova16} based on \omegas$=(R_{\rm i}/R_{\rm co})^{3/2}$. 
The median/\MAD\ \omegas\ is \medianOmega$\pm$\MADOmega. 
Four CTTS are in the weak propeller regime ($\omega_s\sim1$), and one is in the strong propeller regime ($\omega_s\gg1$). Six CTTS are stable, eight are unstable chaotic, and the remaining 28 are unstable ordered.
There may be a lower limit on \omegas\ around 0.04, indicated by the dashed grey line on Figure~\ref{fig:RiRco}. This means that the stellar angular velocity is always at least 4\% that of the infalling disk material in these CTTS.

We noted in \paperI\ that the typical assumption that \ri$=$5~\rstar\ has a small impact on the \mdot\ measurement. However, it has a significant impact on the accretion stability regime classification. If we took the 2D distribution in Figure~\ref{fig:RiRco} and collapsed it into a 1D distribution at \ri$=$5~\rstar, we would find very different \omegas\ values. There would be fourteen propellers (four of which would be strong propellers, with $\omega_s>2$), eleven stable, six unstable chaotic, and sixteen unstable ordered. CTTS are not expected to be in the propeller regime for long periods of time, as the star should spin down through propeller-driven outflows and star-disk interaction torques with a spin-down timescale of $\sim0.8-2.7$~Myr \citep[][]{Ustyugova2006,Lii2014,Ireland2022,Zhu2025,Takasao2025}.
Therefore, our measured propeller occurrence of 5/47 is more likely than the 14/47 implied from assuming \ri$=$5~\rstar.


\subsection{Light curve morphology results}   \label{subsec:VariabilityMetrics}

Figure~\ref{fig:QM_violin} presents QM variability metric results for the sample.
As shown in the left panel, the Q metric spans the full expected [0--1] range, with a minimum value of 0.03 in the sector 39 light curve of Sz~130 and a maximum value of 0.98 in the sector 44 light curve of DM~Tau. Therefore, we do not need to apply additional Q scaling to align with the suggestions of \cite{Hillenbrand2022}. We visually inspect the light curves sorted by Q and find that the \cite{Cody18} Q boundaries of 0.11 and 0.85 align well with our observed light curve morphologies in TESS data. We begin to see light curves with significant stochasticity above ${\rm Q}\sim0.6$, in line with the boundary of \cite{Cody14}, but it is at values of ${\rm Q}>0.85$ that all observations are almost purely stochastic. Therefore, we choose 0.11 and 0.85 as the Q boundaries between periodic, quasi-periodic, and stochastic light curves.

Similar to other work in the literature, we find general trends between the QM metrics and system characteristics, but no statistically robust empirical correlations. In Section~\ref{subsec:M}, we discuss trends with the symmetry metric M, and in Section~\ref{subsec:Q}, we discuss trends with the periodicity metric Q.
Additional QM metric results are presented in Appendix~\ref{Appsec:additionalQM}.


\subsubsection{The symmetry metric M} \label{subsec:M}

\begin{figure*}[t]
    \centering
    \includegraphics[trim=0pt 6pt 0pt 0pt, clip, width=0.463\linewidth]{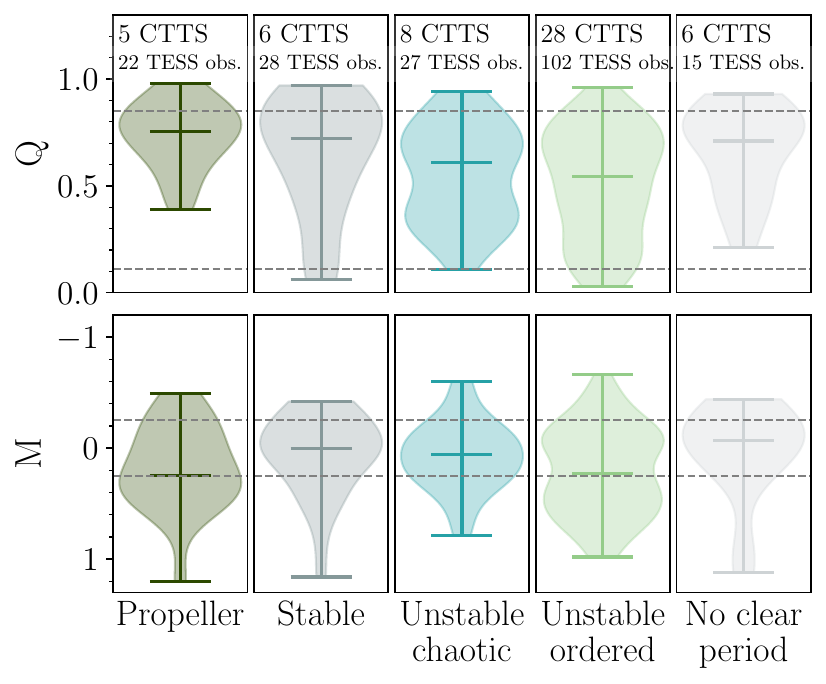} 
    \includegraphics[trim=46pt 7.5pt 0pt 2pt, clip, width=0.53\linewidth]{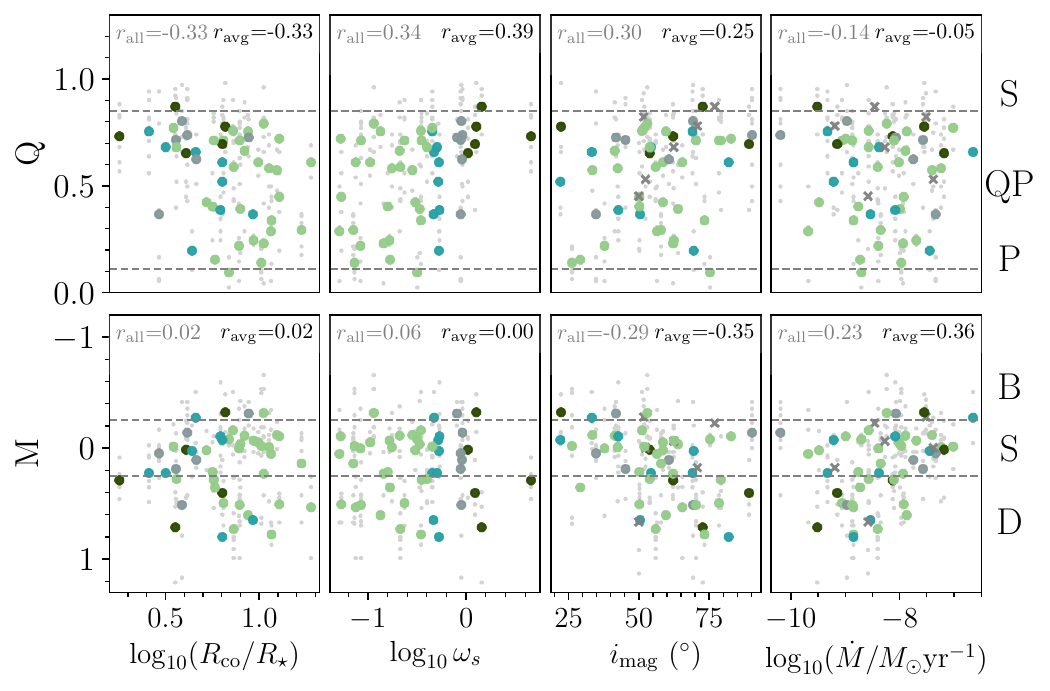}
    \caption{
    QM variability results. Gray dashed lines indicate the divisions between categories as described in Section~\ref{sec:analysis}, with labels on the far right indicating the Q categories (S/Stochastic, QP/Quasi-Periodic, P/Periodic) and M categories (B/Burster, S/Symmetric, D/Dipper).
    \textit{Left}: Violin plots showing the QM distributions grouped by accretion stability regime as determined by \omegas. The horizontal lines indicate the minimum, median, and maximum values, and the width indicates the frequency of that value. The number of unique CTTS in each category, and the total number of TESS observations analyzed, are indicated. \textit{Right}: Correlations between the QM metrics and CTTS properties. Small gray points mark all individual measurements, and colored points mark the average value of Q and M for each CTTS (where the colors correspond to the accretion stability regimes in the left panel). Gray x markers indicate the six CTTS without confident period measurements. The Pearson correlation coefficients are indicated for both sets. None show strong empirical relationships, but there are general trends as discussed in the text.
    }
    \label{fig:QM_violin}
\end{figure*}

Figure~\ref{fig:QM_violin} (left) shows the QM results separated by the predicted accretion stability regimes, and Figure~\ref{fig:QM_violin} (right) shows the average Q and M values for each target as a function of \rco, \imag, \omegas, and \mdot. 
We see that \imag\ is moderately correlated with M with a Pearson correlation coefficient of $r=-0.35$, and $\log_{10}(\dot{M}/M_\odot{\rm yr}^{-1})$ is moderately correlated with M with $r=0.36$ as expected from previous work \citep[][]{Stauffer2014Bursters,Bodman2017,Cody18,Robinson2021}. 
Overall, the light curves in our sample are biased towards being symmetric or dippers, with $M_{\rm med}=0.08$ and $M_{\rm mean}=0.15$ (see Figure~\ref{fig:QM_violin}, left). 

There are 17 CTTS whose average M values fall in the dipper category, and all but one of these have $i_{\rm mag}\gtrsim50^\circ$.
RX~J1852.3-3700 has a lower inclination of 29\degrees, and its single TESS observation has $M=0.35$. However, inspection of the light curve shows that the variability is consistent with rotational modulation by two hotspots separated by approximately one quarter phase. This increases the median flux while leaving the bottom decile of the fluxes unchanged, producing a large M value. Therefore, this is unlikely a physical dipper, but rather a statistical effect.
There are 5 CTTS whose average M values fall in the burster category: CVSO~90, CVSO~109~A, DM~Tau, DG~Tau, and VZ~Cha. These tend towards lower \imag\ (42\degrees, 33\degrees, 22\degrees, 52\degrees, and 53\degrees, respectively) and higher \mdot\ ($\gtrsim6\times10^{-9}$~\msunyr).


\startlongtable
\begin{deluxetable*}{>{\raggedright\arraybackslash}p{2.5cm}p{2cm}>{\raggedright\arraybackslash}p{2.1cm}lllllll}
\tablewidth{\textwidth}
\tabletypesize{\small}
\tablecaption{Sample and Results \label{tab:results}}
\tablehead{
\colhead{Object} & \colhead{TESS Sector(s)} & \colhead{QM Type(s)} & \colhead{\mstar} & \colhead{\imag} & \colhead{$\log_{10}$\mdot} & \colhead{\prot} & \colhead{\rco} & \colhead{\ri} & \colhead{\omegas} \\
\colhead{} & \colhead{} & \colhead{} & \colhead{(\msun)} & \colhead{(\degrees)} & \colhead{[\msunyr]} & \colhead{(days)} & \colhead{(\rstar)} & \colhead{(\rstar)} & \colhead{} \\
\colhead{(1)} & \colhead{(2)} & \colhead{(3)} & \colhead{(4)} & \colhead{(5)} & \colhead{(6)} & \colhead{(7)} & \colhead{(8)} & \colhead{(9)} & \colhead{(10)}
}
\startdata
\multicolumn{10}{c}{\textbf{\propeller{Propeller}}} \\
\hline
2MASS J11432669-7804454 & 11, 12, 38, 39, 64--66 & QPD, QPS, SD & 0.12 & $62\pm10$ & $-8.10\pm0.19$ & ${1.38}^{+0.03}_{-0.02}$ & ${1.79}^{+0.17}_{-0.15}$ & ${4.89}$ & ${4.53}$ \\
SSTc2d J160000.6-422158 & 12, 39, 65 & QPD, SD & 0.15 & $72\pm2$ & $-9.53\pm0.12$ & ${2.20}^{+0.08}_{-0.20}$ & ${3.53}^{+0.37}_{-0.47}$ & ${4.45}^{5.88}_{3.77}$ & ${1.41}^{2.14}_{1.10}$ \\
DM Tau & 43, 44, 70, 71 & QPB, QPS, SB & 0.38 & $22\pm3$ & $-7.69\pm0.00$ & ${7.87}^{+1.09}_{-0.90}$ & ${6.53}^{+1.16}_{-0.95}$ & ${7.57}^{7.90}_{6.80}$ & ${1.25}^{1.33}_{1.06}$ \\
MY Lup & 12, 39, 65 & QPD, QPS, SD & 1.06 & $89\pm0$ & $-9.18\pm0.08$ & ${2.60}^{+0.11}_{-0.10}$ & ${6.29}^{+0.69}_{-0.61}$ & ${7.17}^{7.96}_{4.24}$ & ${1.22}^{1.42}_{0.55}$ \\
CV Cha & 12, 39, 64--66 & QPB, QPD, QPS & 1.40 & $54\pm2$ & $-7.24\pm0.08$ & ${3.65}^{+0.23}_{-0.15}$ & ${4.04}^{+0.51}_{-0.40}$ & ${4.14}^{4.19}_{4.00}$ & ${1.04}^{1.06}_{0.99}$ \\
\hline \multicolumn{10}{c}{\textbf{\stable{Stable}}} \\ \hline 
RECX 16 & 11--13, 64--66 & QPB, QPS, SD & 0.05 & $90\pm0$ & $-10.18\pm0.16$ & ${1.43}^{+0.02}_{-0.03}$ & ${4.09}^{+0.38}_{-0.35}$ & ${3.83}^{4.16}_{3.22}$ & ${0.91}^{1.02}_{0.70}$ \\
CVSO 58 & 6, 32 & QPS & 0.81 & $60\pm2$ & $-7.91\pm0.10$ & ${2.84}^{+0.21}_{-0.07}$ & ${4.57}^{+0.61}_{-0.41}$ & ${4.24}^{4.63}_{3.88}$ & ${0.89}^{1.02}_{0.78}$ \\
IN Cha & 12, 39, 64, 65 & QPB, QPD, SD & 0.15 & $69\pm3$ & $-9.00\pm0.09$ & ${5.40}^{+0.55}_{-0.38}$ & ${3.82}^{+0.58}_{-0.45}$ & ${3.53}^{3.63}_{3.16}$ & ${0.89}^{0.92}_{0.75}$ \\
Sz 19 & 11, 12, 38, 39, 64, 66, 93 & PS, QPD, QPS, SS & 2.37 & $35\pm3$ & $-7.37\pm0.08$ & ${2.43}^{+0.05}_{-0.04}$ & ${2.89}^{+0.28}_{-0.24}$ & ${2.64}^{4.13}_{2.48}$ & ${0.87}^{1.71}_{0.80}$ \\
DE Tau & 19, 43, 44, 70, 71 & QPD, QPS, SD, SS & 0.35 & $45\pm2$ & $-7.68\pm0.05$ & ${5.93}^{+0.79}_{-0.46}$ & ${3.55}^{+0.62}_{-0.43}$ & ${3.23}^{3.39}_{3.01}$ & ${0.87}^{0.94}_{0.78}$ \\
CVSO 90 & 6, 32 & QPB, QPS & 0.62 & $42\pm1$ & $-8.22\pm0.04$ & ${5.03}^{+0.67}_{-0.39}$ & ${8.71}^{+1.51}_{-1.06}$ & ${7.49}^{7.75}_{7.46}$ & ${0.80}^{0.84}_{0.79}$ \\
\hline \multicolumn{10}{c}{\textbf{\unstablechaotic{Unstable chaotic}}} \\ \hline 
CS Cha & 11, 12, 38, 39, 64--66, 93 & QPB, QPS & 1.22 & $43\pm2$ & $-8.10\pm0.10$ & ${4.94}^{+0.14}_{-0.11}$ & ${6.15}^{+0.62}_{-0.54}$ & ${4.04}^{6.27}_{3.65}$ & ${0.53}^{1.03}_{0.46}$ \\
Sz 117 & 65 & QPD & 0.29 & $82\pm1$ & $-8.80\pm0.08$ & ${4.01}^{+0.32}_{-0.29}$ & ${6.28}^{+0.85}_{-0.74}$ & ${4.09}^{5.16}_{3.05}$ & ${0.53}^{0.74}_{0.34}$ \\
Sz 68 & 38, 65 & QPD, QPS & 1.83 & $69\pm3$ & $-7.42\pm0.06$ & ${4.27}^{+0.35}_{-0.23}$ & ${4.34}^{+0.60}_{-0.47}$ & ${2.82}^{5.19}_{2.56}$ & ${0.52}^{1.31}_{0.45}$ \\
Sz 69 & 12, 38, 65 & QPS & 0.15 & $22\pm15$ & $-8.96\pm0.10$ & ${2.75}^{+0.15}_{-0.09}$ & ${6.27}^{+0.75}_{-0.58}$ & ${4.05}^{9.84}_{3.95}$ & ${0.52}^{1.96}_{0.50}$ \\
CVSO 176 & 6 & QPS & 0.25 & $54\pm2$ & $-8.43\pm0.05$ & ${3.59}^{+0.26}_{-0.22}$ & ${3.14}^{+0.41}_{-0.35}$ & ${1.99}^{2.02}_{1.94}$ & ${0.50}^{0.52}_{0.48}$ \\
CVSO 109 A & 6, 32 & QPB, QPS & 0.50 & $33\pm5$ & $-6.83\pm0.07$ & ${6.52}^{+1.39}_{-0.69}$ & ${4.54}^{+1.04}_{-0.64}$ & ${2.73}^{6.48}_{2.40}$ & ${0.47}^{1.71}_{0.39}$ \\
RX J1842.9-3532 & 13 & QPD & 1.04 & $50\pm1$ & $-8.46\pm0.03$ & ${2.54}^{+0.10}_{-0.07}$ & ${9.12}^{+0.98}_{-0.84}$ & ${5.47}^{5.85}_{4.32}$ & ${0.46}^{0.51}_{0.33}$ \\
RECX 9 & 11--13, 37--39, 64--66 & QPD, QPS, SS & 0.12 & $69\pm2$ & $-9.32\pm0.05$ & ${1.95}^{+0.04}_{-0.08}$ & ${2.55}^{+0.24}_{-0.26}$ & ${1.50}^{1.53}_{1.44}$ & ${0.45}^{0.46}_{0.42}$ \\
\hline \multicolumn{10}{c}{\textbf{\unstableordered{Unstable ordered}}} \\ \hline 
TX Ori & 6, 32 & QPS, SS & 1.09 & $52\pm5$ & $-7.01\pm0.13$ & ${4.63}^{+0.48}_{-0.26}$ & ${3.45}^{+0.53}_{-0.38}$ & ${2.02}^{2.19}_{1.80}$ & ${0.45}^{0.51}_{0.38}$ \\
Sz 10 & 11, 12, 38, 39, 64, 65 & QPD, QPS & 0.17 & $54\pm0$ & $-8.51\pm0.04$ & ${3.60}^{+0.12}_{-0.24}$ & ${3.58}^{+0.37}_{-0.41}$ & ${1.93}^{1.97}_{1.79}$ & ${0.40}^{0.41}_{0.35}$ \\
UX Tau A & 43, 44, 70, 71 & QPD, QPS & 1.50 & $79\pm2$ & $-8.17\pm0.09$ & ${3.70}^{+0.34}_{-0.15}$ & ${5.65}^{+0.81}_{-0.56}$ & ${2.81}^{6.03}_{2.46}$ & ${0.35}^{1.10}_{0.29}$ \\
SSTc2d J161243.8-381503 & 12, 39, 65 & QPD & 0.57 & $73\pm2$ & $-8.82\pm0.10$ & ${8.40}^{+2.37}_{-0.77}$ & ${11.44}^{+3.14}_{-1.51}$ & ${5.61}^{7.08}_{3.36}$ & ${0.34}^{0.49}_{0.16}$ \\
Sz 72 & 12, 38, 65 & QPB, QPS, SS & 0.37 & $51\pm0$ & $-8.49\pm0.02$ & ${4.80}^{+1.22}_{-0.23}$ & ${7.16}^{+1.84}_{-0.75}$ & ${3.50}^{3.90}_{3.14}$ & ${0.34}^{0.40}_{0.29}$ \\
RX J1556.1-3655 & 12, 65 & QPB, QPD & 0.50 & $64\pm6$ & $-8.03\pm0.21$ & ${5.80}^{+0.59}_{-0.45}$ & ${7.81}^{+1.19}_{-0.95}$ & ${3.78}^{3.80}_{2.75}$ & ${0.34}^{0.34}_{0.21}$ \\
Sz 130 & 12, 39, 65 & PS, QPS & 0.24 & $75\pm2$ & $-8.68\pm0.07$ & ${6.30}^{+0.64}_{-0.54}$ & ${6.81}^{+1.03}_{-0.87}$ & ${3.16}^{3.30}_{2.60}$ & ${0.32}^{0.34}_{0.24}$ \\
SSTc2d J161344.1-373646 & 12, 39 & QPS & 0.13 & $58\pm1$ & $-9.47\pm0.04$ & ${1.86}^{+0.06}_{-0.04}$ & ${5.20}^{+0.54}_{-0.45}$ & ${2.34}^{2.58}_{1.97}$ & ${0.30}^{0.35}_{0.23}$ \\
V510 Ori & 6, 32 & QPS & 0.76 & $33\pm0$ & $-7.63\pm0.03$ & ${9.40}^{+1.41}_{-1.64}$ & ${12.31}^{+2.28}_{-2.28}$ & ${5.29}^{6.41}_{5.00}$ & ${0.28}^{0.38}_{0.26}$ \\
Hn 5 & 11, 12, 39, 64, 65 & QPD, SD & 0.15 & $78\pm1$ & $-8.95\pm0.02$ & ${7.93}^{+0.77}_{-0.46}$ & ${6.35}^{+0.95}_{-0.70}$ & ${2.43}^{2.50}_{1.76}$ & ${0.24}^{0.25}_{0.15}$ \\
SZ Cha & 11, 12, 38, 39, 64--66, 93 & QPD & 1.22 & $56\pm3$ & $-8.34\pm0.13$ & ${3.20}^{+0.21}_{-0.18}$ & ${7.23}^{+0.91}_{-0.79}$ & ${2.58}^{4.16}_{2.08}$ & ${0.21}^{0.44}_{0.15}$ \\
WZ Cha & 12, 39, 64, 65 & QPS, SB & 0.29 & $41\pm0$ & $-7.79\pm0.05$ & ${5.30}^{+0.64}_{-0.25}$ & ${8.22}^{+1.36}_{-0.85}$ & ${2.89}^{3.33}_{2.86}$ & ${0.21}^{0.26}_{0.21}$ \\
AA Tau & 43, 44, 70, 71 & QPB, QPD, QPS & 0.80 & $42\pm2$ & $-7.11\pm0.07$ & ${8.38}^{+1.87}_{-0.81}$ & ${11.11}^{+2.62}_{-1.50}$ & ${3.47}^{6.18}_{3.36}$ & ${0.17}^{0.42}_{0.17}$ \\
RX J1852.3-3700 & 67 & QPD & 0.82 & $29\pm1$ & $-8.71\pm0.06$ & ${2.83}^{+0.13}_{-0.09}$ & ${5.75}^{+0.64}_{-0.54}$ & ${1.75}^{1.91}_{1.58}$ & ${0.17}^{0.19}_{0.14}$ \\
V505 Ori & 6, 32 & QPS & 0.81 & $62\pm2$ & $-7.88\pm0.06$ & ${7.15}^{+1.13}_{-0.64}$ & ${9.21}^{+1.76}_{-1.19}$ & ${2.78}^{3.91}_{2.48}$ & ${0.17}^{0.28}_{0.14}$ \\
Sz 71 & 12, 38, 65 & QPD, QPS & 0.37 & $50\pm3$ & $-8.67\pm0.08$ & ${4.13}^{+0.34}_{-0.24}$ & ${5.59}^{+0.77}_{-0.62}$ & ${1.64}^{1.96}_{1.56}$ & ${0.16}^{0.21}_{0.15}$ \\
Sz 82 & 12, 65 & PD, QPS & 0.68 & $62\pm1$ & $-7.82\pm0.07$ & ${7.23}^{+1.10}_{-0.52}$ & ${10.40}^{+1.94}_{-1.23}$ & ${2.86}^{3.56}_{2.27}$ & ${0.14}^{0.20}_{0.10}$ \\
CVSO 146 & 6, 32 & QPD & 0.86 & $60\pm1$ & $-7.78\pm0.03$ & ${5.50}^{+0.62}_{-0.47}$ & ${8.60}^{+1.37}_{-1.09}$ & ${2.25}^{2.35}_{2.11}$ & ${0.13}^{0.14}_{0.12}$ \\
VZ Cha & 12, 37, 39, 64, 65 & QPB, QPS, SB & 0.64 & $53\pm2$ & $-8.02\pm0.03$ & ${5.20}^{+0.48}_{-0.38}$ & ${10.47}^{+1.53}_{-1.25}$ & ${2.47}^{2.97}_{2.22}$ & ${0.11}^{0.15}_{0.10}$ \\
CVSO 107 & 6, 32 & QPS & 0.53 & $58\pm1$ & $-7.97\pm0.07$ & ${6.40}^{+2.06}_{-0.56}$ & ${9.78}^{+2.94}_{-1.26}$ & ${2.19}^{2.41}_{1.73}$ & ${0.11}^{0.12}_{0.07}$ \\
RECX 11 & 11--13, 37--39, 64--66 & QPD, QPS & 0.78 & $70\pm2$ & $-8.80\pm0.04$ & ${4.90}^{+0.52}_{-0.47}$ & ${7.76}^{+1.20}_{-1.04}$ & ${1.50}^{2.49}_{1.45}$ & ${0.09}^{0.18}_{0.08}$ \\
DN Tau & 43, 44, 70, 71 & PS, QPB, QPD & 0.58 & $38\pm2$ & $-8.32\pm0.09$ & ${6.32}^{+0.94}_{-0.36}$ & ${7.69}^{+1.42}_{-0.84}$ & ${1.47}^{1.91}_{1.39}$ & ${0.08}^{0.12}_{0.08}$ \\
Sz 111 & 12, 39, 65 & QPD & 0.58 & $65\pm0$ & $-8.80\pm0.00$ & ${11.88}^{+4.18}_{-2.14}$ & ${18.65}^{+5.97}_{-3.51}$ & ${3.33}^{4.49}_{3.24}$ & ${0.08}^{0.12}_{0.07}$ \\
SY Cha & 11, 12, 38, 39, 64--66 & PB, PS, QPB, QPD, QPS & 0.70 & $26\pm4$ & $-7.85\pm0.11$ & ${6.12}^{+0.77}_{-0.37}$ & ${10.16}^{+1.72}_{-1.14}$ & ${1.79}^{1.84}_{1.64}$ & ${0.07}^{0.08}_{0.06}$ \\
Sz 45 & 12, 38, 39, 64--66 & QPD, QPS & 0.60 & $58\pm3$ & $-8.22\pm0.07$ & ${13.17}^{+1.54}_{-0.98}$ & ${16.63}^{+2.70}_{-2.00}$ & ${2.86}^{3.07}_{2.39}$ & ${0.07}^{0.08}_{0.05}$ \\
DK Tau A & 43, 44, 70, 71 & QPD & 0.63 & $50\pm2$ & $-7.72\pm0.05$ & ${8.20}^{+1.49}_{-0.72}$ & ${12.65}^{+2.62}_{-1.63}$ & ${1.81}^{2.51}_{1.60}$ & ${0.05}^{0.09}_{0.05}$ \\
RY Lup & 12 & QPS & 1.40 & $82\pm1$ & $-8.68\pm0.05$ & ${3.70}^{+0.34}_{-0.23}$ & ${12.65}^{+1.83}_{-1.42}$ & ${1.78}^{2.81}_{1.51}$ & ${0.05}^{0.10}_{0.04}$ \\
SSTc2d J160830.7-382827 & 12, 39, 65 & PS, QPB, QPD & 1.38 & $56\pm4$ & $-9.58\pm0.08$ & ${6.21}^{+0.84}_{-0.42}$ & ${11.39}^{+1.99}_{-1.32}$ & ${1.58}^{1.80}_{1.56}$ & ${0.05}^{0.06}_{0.05}$ \\
\hline \multicolumn{10}{c}{\textbf{{No clear period}}} \\ \hline 
DG Tau A & 43, 44, 70, 71 & QPB, QPS, SS & 1.40 & $52\pm6$ & $-7.42\pm0.21$ & (6.3) & \dots & ${4.29}$ & \dots \\
Sz 75 & 12, 38, 65 & QPB, QPS & 0.82 & $52\pm1$ & $-7.27\pm0.03$ & (7.8) & \dots & ${2.83}^{3.11}_{2.60}$ & \dots \\
Sz 77 & 12, 38, 65 & QPD, QPS & 0.75 & $50\pm9$ & $-8.58\pm0.27$ & (10.7) & \dots & ${2.44}^{3.27}_{2.44}$ & \dots \\
Sz 84 & 12 & QPS & 0.16 & $70\pm2$ & $-9.19\pm0.10$ & (7.88) & \dots & ${1.90}^{1.96}_{1.84}$ & \dots \\
Sz 110 & 65 & SS & 0.29 & $77\pm2$ & $-8.39\pm0.14$ & (7.05) & \dots & ${1.67}^{2.20}_{1.42}$ & \dots \\
Sz 129 & 12, 39, 65 & QPS & 0.72 & $62\pm2$ & $-8.09\pm0.07$ & (6.75) & \dots & ${3.06}^{5.23}_{2.36}$ & \dots \\
 \enddata
 \tablenotetext{}{Columns: (1) Object name, (2) TESS sectors with usable data as described in Section~\ref{subsec:ProtRco}, (3) unique QM variability classes into which the TESS sectors fall, where the first portion indicates the Q category (P/Periodic, QP/Quasi-Periodic, S/Stochastic), and the second indicates the M category (D/Dipper, S/Symmetric, B/Burster), (4)--(6) stellar mass (which has typical uncertainties of 0.1~dex), magnetospheric inclination, and accretion rate from \cite{Pittman2025}, (7) Period, with uncertainties, found in periodogram analysis, (8) Corotation radius with uncertainties, assuming \prot\ to be the stellar rotation period, (9) Magnetospheric truncation radius from \paperI, with the upper and lower values indicating the maximum and minimum \ri\ values (not uncertainties) found across the multi-epoch accretion flow modeling, and (10) The fastness parameter found from \rco\ and \ri, with the upper and lower values again showing the range found in the multi-epoch modeling. 
 When there is no range for \ri\ or \omegas, that CTTS had only one epoch of observation.
 Objects are sorted by \omegas\ and grouped by the predicted accretion stability regime. For the 6 CTTS with good TESS data but no clear period, we calculate the Q metric by assuming a possible period that results in a classification of quasi-periodic in at least one TESS sector. These low-confidence periods are shown in parentheses.
 }
\end{deluxetable*}

\subsubsection{The periodicity metric Q} \label{subsec:Q}

The hydrodynamic models and associated synthetic light curves of \cite{Robinson2021} indicate that periodicity should increase moderately with \rco. These models make the simplifying assumption that \omegas\ is fixed at 0.35, which happens to be consistent with our median measured \omegas\ of \medianOmega$\pm$\MADOmega. Our results are in broad agreement, with a correlation coefficient of $r=-0.33$ between \rco\ and Q. Systems with $R_{\rm co}\leq5$~\rstar\ have a median Q of 0.70, whereas those with $R_{\rm co}>5$~\rstar\ have a median Q of 0.53.
Propellers tend to have smaller \rco\ values, and they have a Q distribution that is comparable to the six CTTS that lack any clear period determination (whose Q values were determined using the tentative periods in Table~\ref{tab:results}). This indicates a chaotic accretion/occultation configuration that may be due in part to the small \rco\ values (i.e., fast stellar rotation).

\section{Discussion} \label{sec:discussion}

Here we explore the implications of our \ri, \rco, \omegas, and QM metric results for angular momentum transport and the formation of short period exoplanets.
Section~\ref{subsec:equilibrium} discusses rotational equilibrium and the influence of magnetospheric outflows (Section~\ref{subsec:RiRcoWinds}), episodic accretion (Section~\ref{subsec:EA}), and accretion-powered stellar winds (Section~\ref{subsec:APSWs}). Then, Sections~\ref{subsec:RiVar_long} and \ref{subsec:RiVar_short} describe observational evidence of long- and short-timescale variability of accretion stability regimes. Section~\ref{subsec:QMsignificance} presents implications of the QM results.
Finally, Section~\ref{subsec:exoplanets} compares our magnetospheric results with the observed locations of ultra-short-period exoplanets.

\subsection{Are most CTTS in rotational equilibrium?} \label{subsec:equilibrium}

\cite{Zhu2025} performed 3D MHD simulations of rotating CTTS and found the following empirical relationship between \omegas\ and the spin-up/spin-down torque due to the star-disk interaction, $\dot{J}_{\star}$:  

\begin{equation} \label{eq:Jdot}
    \dot{J}_{\star} = (0.83 - 2.68\omega_s^4)\times\left(\dot{M}\sqrt{GM_\star R_i}\right)
\end{equation}

\noindent The zero-torque solution|that is, the equilibrium spin state|occurs around $\omega_s=0.75$, which falls within the stable accretion regime of \cite{blinova16}.
The median \omegas\ in our sample is \medianOmega$\pm$\MADOmega, in the unstable ordered regime, indicating that most systems are not in an equilibrium spin state.
This is consistent with the results of \cite{thanathibodee23}, who applied the same accretion flow model to a sample of low accretors and found that the majority are in the unstable regime, with about half of their sample in the unstable ordered regime. 

Low accretors are expected to be nearing the end of the CTTS phase. Given that even these evolved systems are in the unstable regime, in which the accreting material has a higher angular momentum than the star, we conclude that CTTS may be slow to reach spin equilibrium, and spin-down processes may be very efficient.
These results further confirm that disk locking is unlikely, as a locked disk would be expected to transfer angular momentum to the star until it reaches equilibrium.
Primary sources of angular momentum loss in CTTS are magnetospheric outflows, episodic accretion, and accretion-powered stellar winds. We find potential evidence for all three in this work, presented in Sections~\ref{subsec:RiRcoWinds}--\ref{subsec:APSWs}.

\subsubsection{Magnetospheric outflows}   \label{subsec:RiRcoWinds}

Magnetospheric outflows are predicted to be important contributors to the angular momentum evolution of CTTS and their inner disks \citep{Shu2000,romanova2018,Gaidos2024}. 
These typically take one of two forms: centrifugally-driven conical winds launched from the disk-magnetosphere boundary at \ri\ \citep[e.g.,][]{romanova2009,Lii2014}, or plasmoids launched by field inflation and magnetic reconnection events caused by the differential rotation of the star and disk, usually referred to as magnetospheric ejections (MEs) or magnetic bubbles \citep[e.g.,][]{ZanniFerreira2013,Ireland2022,Zhu2025}.
In this work, we refer to these as MEs.
Conical winds are expected to be most important for strong propellers, whereas MEs are more significant for slower rotators \citep{romanova2018}.
\cite{romanova2009} and \cite{Takasao2022,Takasao2025} found that conical winds appear in both slow and fast rotators; however, the mass ejection efficiency depends on the rotation, with higher-\omegas\ systems losing more matter to the wind.

Conical winds can appear with accretion bursts or remain steady over longer timescales, whereas MEs are inherently bursty due to the nature of reconnection events.
Both forms of magnetospheric outflows transfer angular momentum between the star and the disk as part of the star-disk interaction.
While conical winds reduce stellar spin-up by carrying away mass and angular momentum from the disk, MEs can result in either spin-up or spin-down depending on the radial extent over which the stellar magnetic field is connected to the disk \citep[specifically with respect to \rco; see][]{Ireland2022}.

To compare the spin states of different CTTS,
we obtain the torque normalized to the stellar angular momentum, $\dot{J}_{\star}/J_\star$, using Equation~\ref{eq:Jdot} and

\begin{equation}  \label{eq:Jstar}
    J_\star = k^2M_\star R_\star^2 \Omega_\star,
\end{equation}

\noindent with $k^2$ taken to be 0.2 \citep[where $k$ is the dimensionless radius of gyration; see][]{ArmitageClarke1996}, and $\Omega_\star=2\pi/P_{\rm rot}$. 
We show $\dot{J}_{\star}/J_\star$ for each CTTS as the colorbar in Figure~\ref{fig:RiRco}. 
Systems close to spin equilibrium are black, those in the spin-down regime are pink, and those in the spin-up regime are green. Clearly, the strong propeller in the sample (2MASSJ11432669-7804454\footnote{Note that if this source is a physical binary, the 0\farcs165 separation corresponds to 31~au at a distance of 190~pc. Future higher-resolution observations should examine whether the companion's contribution has a significant impact on the magnetospheric accretion results presented here.}) is in a strong spin-down regime, with an expected spin-down timescale of 490~years (where the spin-up/spin-down timescale is given by $\tau = J_\star/{\dot{J}_\star}$). 
The majority of the sample is in the spin-up regime. This could indicate either that they have had insufficient time to reach spin equilibrium, or that there are spin-down processes actively preventing them from reaching spin equilibrium.

The 2.5D axisymmetric simulations of \cite{Ireland2022} predict that MEs should exert spin-down torques on the star when $\omega_s\gtrsim0.46$ (which is the boundary between the unstable chaotic and unstable ordered regimes in \citealt{blinova16}). This might explain why the majority of CTTS in our sample have $\omega_s<0.46$, as the ME spin-down effect will increase $\tau$ as \omegas\ increases.
We note, though, that the non-axisymmetry of MEs revealed by 3D models \citep[e.g.,][]{Takasao2025,Zhu2025} might reduce their spin-down efficiency, so more work is needed to confirm the critical \omegas\ value that divides whether MEs produce spin-up or spin-down.

The strength of conical wind outflows is predicted to increase with \omegas\ \citep[][]{Ustyugova2006,romanova2009,romanova2018, Lii2014, Zhu2025}.
These winds are expected to have opening angles $\Theta_{\rm wind}\sim40^\circ-60^\circ$, with the specific value depending on \omegas.
Because these winds are launched from \ri, they will cross our line of sight towards the magnetospheric flow if our viewing inclination is $\gtrsim\Theta_{\rm wind}$. 
This will produce blueshifted absorption features in \halpha, which have been seen in hybrid modeling of the magnetosphere and disk winds \citep{Kurosawa2006,Kurosawa2011} as well as in 49\% of the observed \halpha\ profiles in \cite{Reipurth1996}. 
The magnetospheric viewing inclinations (\imag) are above 45\degrees\ for 35/47 CTTS in our sample \citep{Pittman2025}, so we can expect many CTTS to show blueshifted absorption if these conical winds are present.
In this sample, \imag\ and \omegas\ show no correlation, so any correlation between conical wind signatures and \omegas\ should be related to the actual presence of conical winds rather than visibility biases.

\begin{figure}[t]
    \centering
    \includegraphics[width=\linewidth]{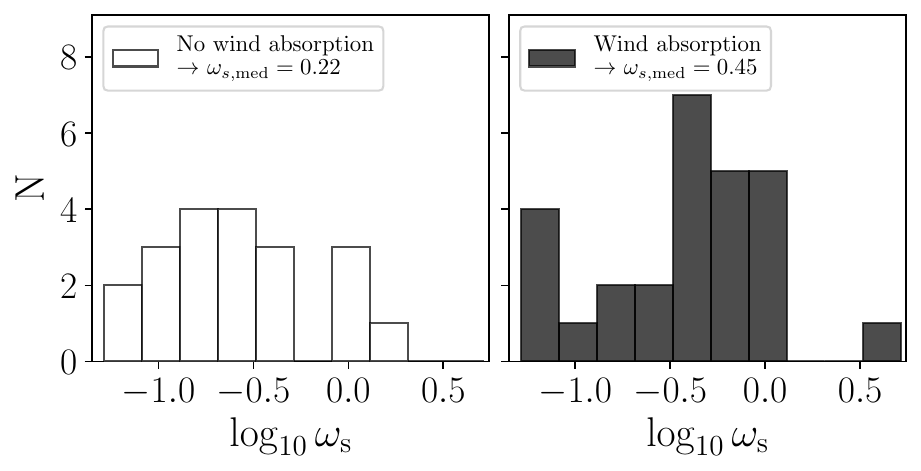}
    \caption{
    Histogram of the fastness parameter for targets that lack (left) or show (right) wind absorption.
    As expected from 3D MHD models, CTTS with higher \omegas\ are more likely to show signatures of conical disk winds.
    }
    \label{fig:omegas}
\end{figure}

Looking at the complete \cite{Pittman2025} sample (including those without \omegas\ measurements), 48\% of these CTTS show blueshifted absorption in all available epochs of observation; 9\% show absorption in only some epochs; and 43\% show no sign of wind absorption.
These occurrence rates are in good agreement with those found by \cite{Reipurth1996}, which includes many of the same CTTS studied here.
Looking at the subset with \omegas\ measurements, we tentatively confirm the expected correlation between outflows and \omegas. CTTS that never show blueshifted absorption have a median \omegas\ of \omegasMed$=$\medianOmegaNoWind,
whereas those with wind absorption have \omegasMed$=$\medianOmegaWind\ (see Figure~\ref{fig:omegas}). 
A larger sample is needed to robustly confirm that the two distributions are distinct, and we will pursue this in future work.
We also find that the detection of a high-velocity component (HVC) in [\ion{O}{1}] 6300~\AA\ indicative of jet emission may be correlated with \omegas, as \omegasMed\ for those with an HVC is 0.48, compared to 0.32 in CTTS without an HVC (J. Campbell-White et al., in preparation).  
Therefore, as \omegas\ increases, we can expect outflow strength to increase, even when the system is not in the propeller regime.

\subsubsection{Episodic accretion}   \label{subsec:EA}

An episodic accretion event begins with the accumulation of material at \ri. This results from the partial or total inhibition of accretion onto the star by the centrifugal barrier in systems with \ri$\sim$\rco\ \citep[][]{SpruitTaam1993,DangeloSpruit2010,DangeloSpruit2012}. This buildup of matter compresses the magnetosphere and eventually breaches the centrifugal barrier and accretes rapidly onto the star. This is followed by a re-expansion of the magnetosphere and an associated outflow of material \citep[for more details, see][]{Lii2014}.
Episodic accretion can occur 
quasi-periodically due to an oscillatory relationship between the centrifugal barrier from stellar rotation and the inner disk ram pressure \citep[][]{Lii2014}, or stochastically due to intrinsic inhomogeneities in the inner disk \citep[e.g.,][]{Stauffer2014Bursters,re19}. Much of the literature focuses on episodic accretion bursts of large magnitude \citep[e.g., EX~Lup,][]{Wang2023,CruzSaenzdeMiera2023,Pouilly2024}. However, it likely occurs on smaller scales as well, as has been proposed for GM~Aur \citep{Bouvier2023}. Here, we find one likely low-magnitude episodic accretion event in CS~Cha with the potential for angular momentum loss. 

CS~Cha is a single-lined spectroscopic binary with a separation of 0\farcs03 ($\sim$5~au), an $R$-band flux ratio of 3.2, and a circumbinary disk \citep{Ginski2024}. It has an orbital period of at least 2482~days \citep[][]{Guenther2007}, so short-timescale accretion modulation by orbital dynamics is not expected. While the disk is viewed at low inclination \citep[$i_{\rm disk}\sim20^\circ$,][]{Ginski2024}, \imag\ appears to be at least 35\degrees\ \citep{Pittman2025}.
This indicates obliquity between the disk and the magnetosphere.

CS~Cha shows blueshifted absorption indicative of winds in only one epoch of observation, and this epoch coincides with an increase in \ri\ (and therefore in \omegas). \paperI\ masked out regions with blueshifted absorption when fitting the flow models, whereas here we include a Gaussian absorption component to take the wind into account. 
CS~Cha was observed three times with UVES (2022-05-11, 2022-05-12, and 2022-05-16), which covers exactly one stellar rotation period of 4.94~days. 
The spectra were taken at three different position angles, but they show very weak spectro-astrometric offsets (less than 0\farcs02) that most likely result from the system's multiplicity \citep{Sperling2024}. These do not affect the integrated line profiles.

\begin{figure*}[t]
    \centering
    \includegraphics[width=0.7\textwidth]{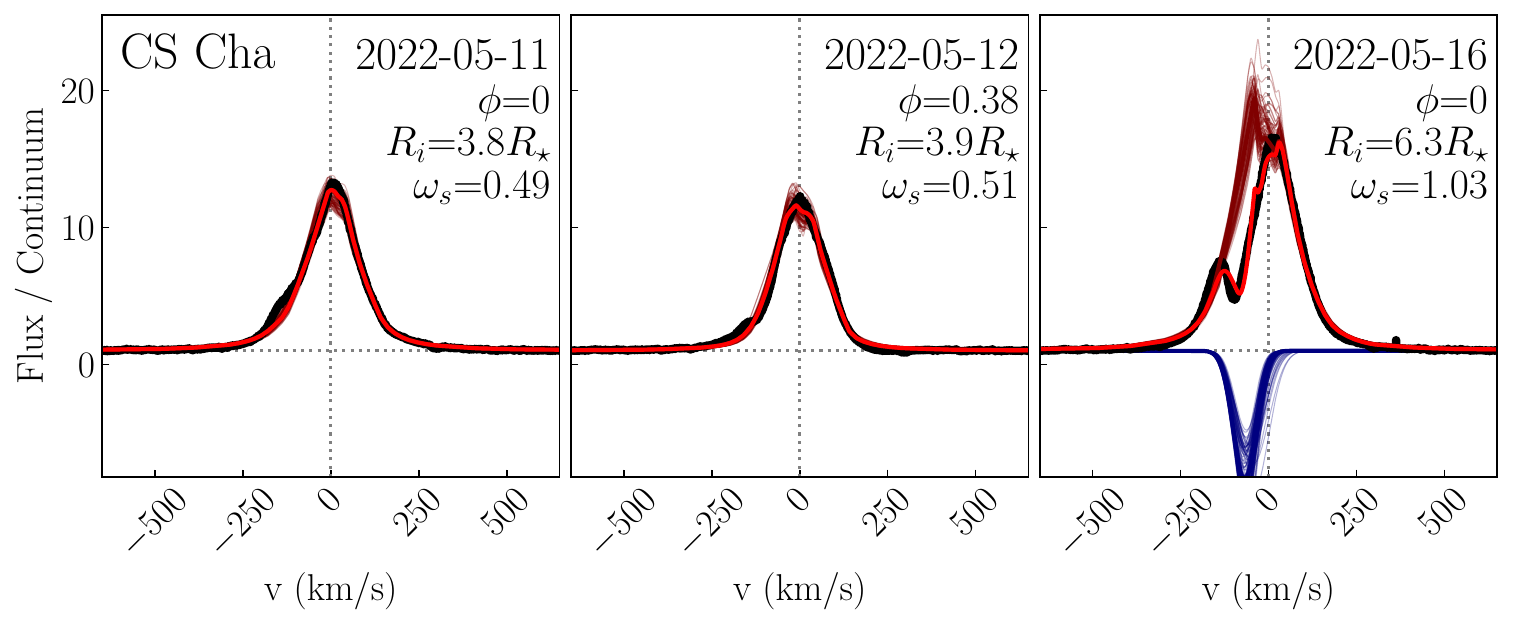}
    \includegraphics[width=0.9\textwidth]{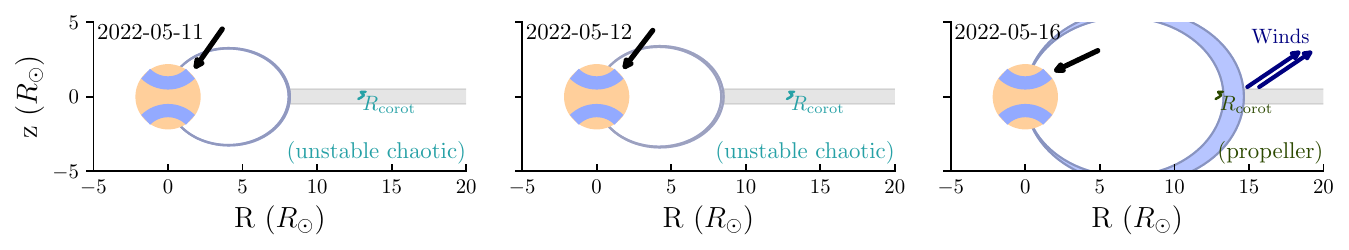}
    \caption{
    \textit{Top}: CS Cha \halpha\ profiles (black) with the top 100 flow models (maroon) and associated wind absorption components (blue). The final weighted-mean model flux is in bright red. The observation date, rotational phase, and associated best-fit \ri\ and \omegas\ values are indicated.
    \textit{Bottom}: Visual representation of the best-fit accretion flow configurations for the same epochs as above. The appearance of strong blueshifted absorption from winds is accompanied by an increase of \ri, pushing the system into the propeller regime. The black arrows indicate our best-fit flow viewing inclination, \imag.
    }
    \label{fig:cscha-variability}
\end{figure*}

As shown in Figure~\ref{fig:cscha-variability}, the \halpha\ profiles on the first two nights are similar, showing only weak signs of blueshifted wind absorption. They are best fit with \ri$\sim$3.9~\rstar, corresponding to \omegas$\sim$0.5 in the unstable chaotic accretion regime. 
However, in the final observation, the best-fit \ri\ increases to 6.3~\rstar, and the system enters the propeller regime with \omegas$=$1.03. This is accompanied by a 70\% decrease in \mdot\ (from $1.11\times10^{-8}$ to $0.65\times10^{-8}$~\msunyr), an increase in \imag\ from 37\degrees\ to 65\degrees\ \citep[potentially indicative of polar accretion along higher-latitude open field lines, which is expected in the propeller regime;][]{romanova2018,Zhu2025}, a factor of 9 increase in the flow width \rw, and very strong blueshifted absorption indicative of an outflow with a projected velocity of $v_{\rm LOS}=-67$~km/s. 
Because the first and last observations occurred at the same phase in stellar rotation, any changes between them should be temporal rather than geometrical effects.

\cite{romanova2018} provide relations for the time-averaged maximum velocity ($\langle v_{\rm max}\rangle$) and opening angle ($\langle\Theta_{\rm wind}\rangle$) of conical winds as a function of \omegas\ for CTTS in the propeller regime. In the relatively large magnetosphere of CS~Cha (6.3~\rstar), the relations are $\langle v_{\rm max}/v_{\rm esc}\rangle=0.16e^{0.61\omega_s}$ and $\langle\Theta_{\rm wind}\rangle=56.2\omega_s^{-0.07}$. For $\omega_s=1.03$ and $v_{\rm esc}(r=R{\rm_i})=187$~km/s, this gives $\langle v_{\rm max}\rangle=56$~km/s and $\langle\Theta_{\rm wind}\rangle=56^\circ$. We can then project $\langle v_{\rm max}\rangle$ to the line of sight according to $\cos({i_{\rm mag}-\Theta_{\rm wind}})$ across the range of \imag\ found for CS~Cha (35--65\degrees). This gives a final expected $v_{\rm LOS}$ between $-53$~km/s to $-56$~km/s. Given that the model 
velocities oscillate around $\langle v_{\rm max}/v_{\rm esc}\rangle$ with an amplitude of a factor of two \citep{romanova2018},
this expected $v_{\rm LOS}$ is consistent with the observed wind signature centered at $-67$~km/s with a velocity dispersion of 33~km/s (see Figure~\ref{fig:cscha-variability}, right). 

All spectral changes in CS~Cha between the first and last UVES observation are consistent with a low-magnitude episodic accretion cycle caused by \mdot\ variability \citep[][]{Lii2014,romanova2018,Ireland2022}. 
In this context, the first two UVES observations trace the system while the magnetosphere is compressed and accretion is stronger. 
Then, the last UVES observation occurs when \mdot\ has decreased, the ram pressure on the magnetosphere has weakened, and the magnetosphere has re-expanded.
This produces outflows and carries away angular momentum.
There are too few data to determine whether this process is cyclic in CS~Cha or a single random compression/expansion event. However, the TESS light curves of CS~Cha consistently show flux bursts on top of the rotational modulation (see, e.g., Figure~\ref{fig:TESS_short}). These bursts last between 0.1--0.8~days and may originate from low magnitude episodic compression/accretion/re-expansion cycles. 

The accretion flow model assumes global changes in the magnetosphere, but the compression and re-expansion may not be axisymmetric. This might explain the rarity of these observations, given that there would be both temporal and spatial variability affecting the observability of this phenomenon. Additionally, CS~Cha is classified in the weak propeller regime, which is not expected to produce strong outflows. A localized event along the line of sight may be more likely than a large-scale inflation of the magnetosphere, as field inflation is expected to act locally rather than globally \citep{Zhu2025,Takasao2025}.

\subsubsection{Accretion-powered stellar winds}   \label{subsec:APSWs}
Accretion-powered stellar winds (APSWs) are magnetized stellar winds driven along open magnetic field lines \citep[][]{MattPudritz2005,MattPudritz2008a,MattPudritz2008b}.
They are theorized to be driven by accretion energy, and they can spin down the star efficiently by forcing matter at large distances from the stellar surface to corotate with the star.
The footprints of these open field lines are at high stellar latitudes, so these outflows are unlikely to intercept the magnetospheric accretion flow along the line of sight. Thus, they are unlikely to produce the blueshifted absorption features that were discussed in the previous sections in the case of conical winds launched from \ri.

\cite{ZanniFerreira2011} tested whether observed accretion luminosities (\lacc) and stellar parameters (\mstar, \rstar, \prot, and magnetic field strength $B_\star$) are consistent with spin equilibrium solutions via APSWs.
They found that APSWs are unlikely to account for the majority of angular momentum loss in CTTS, as they require very strong magnetic fields and high mass ejection efficiencies ($f_M$).
We replicated the method of \cite{ZanniFerreira2011} to test whether any CTTS in our sample are compatible with the zero-torque condition due to APSWs, using the following relation to estimate the equatorial magnetic field strengths:

\begin{equation} \label{Eq:B}
        B_{\rm \star,eq} = (2G/C^7)^{1/4}
        \dot{M}^{1/2} M_\star^{1/4} R_i^{7/4} R_\star^{-3}
\end{equation}
\\
\noindent \citep[where $G$ is the gravitational constant and $C$ is taken to be 0.75;][]{Bessolaz2008,ZanniFerreira2011}. Note that the equatorial magnetic field strength is half the polar magnetic field strength for a pure dipole, \Bpolar\ (i.e., $B_{\rm \star,eq} = 0.5B_{\rm \star,polar}$).

This test showed that CTTS in our sample with \omegas$>$0.8 have sufficiently low \lacc\ and high \Beq\ to permit spin equilibrium due to APSWs. However, in agreement with \cite{ZanniFerreira2011}, they require very high ejection efficiencies ($f_M=20-50\%$). Additionally, spin-down due to the star-disk interaction is expected to be efficient at these high \omegas\ values, potentially removing the need for strong APSW contributions. However, we find that all CTTS in our sample are compatible with having APSWs that carry away a small amount of the accreted matter ($f_M=1-2\%$, which should produce a spin-down torque corresponding to 20--30\% of the accretion torque; see \citealt{ZanniFerreira2013}). Decreasing \mdot\ by 30\% in Equation~\ref{eq:Jdot} can approximately incorporate the effect of an APSW, and this would decrease the torque by 30\%. Such a decrease is insufficient for placing our CTTS in equilibrium states, reinforcing the result that most CTTS are not in spin equilibrium.

\cite{Gehrig2025} find that the distribution of \ri/\rco\ (or, equivalently, \omegas) can distinguish whether APSWs contribute significantly to the spin-down of CTTS. Their stellar population model, which assumes the systems are near an equilibrium torque state, indicates that the wind ($f_M=1\%$) and no-wind ($f_M=0\%$) scenarios can be distinguished given a sample of at least 250 CTTS with \ri\ and \rco\ measurements. While our work contains only 47 CTTS, we will place our results in context by comparing our measurements with their predictions. 
\cite{Gehrig2025} predict that CTTS with significant torque contributions from APSWs will have a mean \omegas\ of \omegasMean$\sim$0.46--0.59. Conversely, those without APSWs will have \omegasMean$\sim$0.72--0.85. In our work, we find \omegasMean$=$\meanOmega\ and \omegasMed$=$\medianOmega\, which points towards the presence of APSWs.

It is important to note that the \omegas\ ranges predicted for the two wind scenarios apply to population-level \omegas\ measurements, whereas the \omegas\ values measured for individual systems are instantaneous and may trace non-equilibrium states.
To filter out the most likely non-equilibrium systems, we first estimate the mass ejection efficiency required to support spin equilibrium at the observed stellar \prot\ and \ri\ using the 2.5D MHD result in \citet[][Equation C1]{Ireland2021}, which incorporates torques from accretion, MEs, and stellar winds. We then remove the systems whose required $f_M$ value is too high to be consistent with the observations based on our above APSW modeling following \cite{ZanniFerreira2011}. 
This filter removes 21/47 CTTS, including all of those with $\omega_s<0.16$.
Considering only the CTTS that are potentially in equilibrium, we find \omegasMean$=$0.60 and \omegasMed$=$0.51, still in better agreement with the APSW scenario.
Thus, a contribution from APSWs is consistent with our observations, but we cannot conclude for certain based on this sample size.

\subsection{Long-timescale accretion stability regime variability driven by large-scale magnetic field changes} \label{subsec:RiVar_long}
 
Accretion stability regime variability has been observed photometrically on timescales of months \citep[][]{Cody2022} to years \citep[][]{McGinnis2015,Sousa2016,Siwak2018,Lin2023}, some of which has been attributed to changes in the intrinsic magnetic topology \citep[e.g.,][]{Sousa2021}. 
One potential example in our sample is Sz~19, which transitions from a highly sinusoidal light curve in 2019 and 2021 to a primarily stochastic morphology in 2023 (see Appendix~\ref{Appsec:MorphologyChanges} for further discussion).
Another example in our sample is AA~Tau, the prototypical dipper \citep[][]{Bouvier1999}. After maintaining a consistent $V$-band magnitude for over 20 years \citep{Grankin2007}, AA~Tau underwent a dimming event of 2~mag in 2011 \citep[][]{Bouvier2013} and has since decreased by a further $\sim$1~mag \citep[K. Grankin, private communication;][]{Bouvier2017}. AA~Tau's dipper morphology has been attributed to a disk warp caused by the system being near the propeller regime (that is, \ri$\sim$\rco) with obliquity between its stellar and magnetic axes \citep{Donati2010,Romanova2013}. While the light curves of AA~Tau have historically exhibited periodic dips at a period of around 8.22~days \citep{Bouvier2007AATau}, TESS sector 43 was dominated by bursts (see Figure~\ref{fig:AATau_TESS} in Appendix~\ref{Appsec:MorphologyChanges}). TESS sectors 44, 70, and 71 show stronger dips, but bursts remain common. 

AA~Tau was observed with VLT/ESPRESSO and VLT/X-Shooter on 2021 December 1--3, less than a month after the end of TESS sector 44, through the PENELLOPE program \citep[Program ID 106.20Z8,][]{manara21}. 
It was also observed with VLT/UVES in late 2008 and early 2009 through Program ID 082.C-0005 (PI A. Scholz), around the same time as the spectropolarimetric observations in \cite{Donati2010} and before AA~Tau's dimming event.
In \cite{Pittman2025}, we analyzed both sets of AA~Tau observations and found that the \halpha\ profiles are dramatically different.
The earlier profiles are double-peaked and reach a maximum flux of ten times that of the continuum. The 2021 observations, conversely, reach over forty times that of the continuum as a result of strong, narrow chromospheric emission \citep[see Figure Set~14 and Table~3 in][]{Pittman2025}. 

Our flow model fits indicated that \imag\ was $59\pm5^\circ$ in 2008/2009, in broad agreement with the minimum magnetic inclination of $50^\circ$ found by \cite{Donati2010}, where the observed inclination increases with stellar rotation due to the $20^\circ$ magnetic obliquity.
Additionally, the 2008/2009 \ri\ value is fairly large at $5.7\pm0.9$~\rstar\ (though still too small to place it in the propeller regime with its \rco\ of 11.1~\rstar).
The 2021 epoch shows a lower \imag\ of $42\pm2^\circ$, smaller than the minimum value determined from the earlier observations of \cite{Donati2010}, and a smaller \ri\ of $3.5\pm0.3$~\rstar. This, combined with the appearance of strong chromospheric emission in \halpha\ and burstiness in the light curves, indicates a large-scale magnetic field change that results in \imag\ and \ri\ variability. Large scale field strength and orientation changes have indeed been observed in CTTS through multi-epoch Zeeman-Doppler imaging \citep[e.g.,][]{Donati2012,Donati2013,Grankin2021,Finociety2023}. 

A January 2025 VLT/ESPRESSO observation of AA~Tau (Program ID 114.27A8; PI J. Campbell-White) shows that the \halpha\ profile continues to show strong chromospheric emission (with a peak flux of forty times that of the continuum, as in December 2021), indicating that the large-scale change may have persisted for at least three years. New magnetic field measurements for AA~Tau would be valuable to determine whether compatible signatures are observed through Zeeman-Doppler imaging techniques.

\subsection{Short-timescale accretion stability regime variability driven by \mdot\ changes} \label{subsec:RiVar_short}

Stability regime variability can also result from changes in \ri\ (or, equivalently, \omegas) driven by accretion rate changes that modify the ram pressure onto the magnetosphere. 3D MHD models predict this form of \ri\ variability on timescales of days, and \ri\ can vary between 4.9-9~\rstar\ for a single simulation run \citep{Lii2014,romanova2018}. 
As shown by the arrows in Figure~\ref{fig:RiRco}, individual objects cross stability regime boundaries over time (assuming that \rco\ is stable on timescales of days to years).

The median time difference (\dt) between observations of a given system in the sample of 66 CTTS from \paperI\ is 2.7~days.
A pairwise comparison of \ri\ measurements for a given CTTS show a median $\Delta R_i$ of 0.06~dex with a \MAD\ of 0.04~dex. The subset of 47 CTTS shown in Figure~\ref{fig:RiRco} has a slightly shorter median \dt\ of 2.0~days and the same median and \MAD\ of $\Delta R_i$. This \ri\ variability is comparable to the measurement uncertainties, as the typical standard deviation for an individual \ri\ measurement is 0.04~dex, and the typical \ri\ model grid spacing is 0.06~dex.
However, the total range of \ri\ measurements for a given CTTS (e.g., $\rm R_{i, max}-R_{i,{\rm min}}$ as shown in Figure~\ref{fig:RiRco}) is larger, with a median of 0.13~dex and \MAD\ of 0.11~dex. This likely comes from real temporal variability. 

\citet[][Figure~2]{romanova2018} shows \ri\ variability of 0.06~dex over 1.5~days and 0.26~dex over 6.7~days, which encompasses the range of $\rm \Delta R_i$ that we observe. 
One of the most conspicuous examples of short-timescale \ri\ variability in our sample is CS~Cha, as we discussed in Section~\ref{subsec:EA}.
Because most of our CTTS have only 4 \halpha\ observations spanning $\sim$3~days, we cannot robustly test for the expected inverse correlation between \ri\ and \mdot\ for a fixed stellar magnetic field. 
However, some CTTS in this sample show hints of such a correlation, and in future work we will pursue dedicated observing campaigns to further quantify accretion-driven \ri\ variability.
We will also examine the effects of rotational modulation of the accretion flow on the inferred variability of magnetospheric properties.


\subsection{Implications of QM results} \label{subsec:QMsignificance}

Dippers are expected to be observed in CTTS with inner disk inclinations above $\sim$60\degrees\ if caused by occultation by a disk warp \citep[e.g.,][]{Bouvier1999,Romanova2013,McGinnis2015,Nagel2019} or above $\sim$45\degrees\ if caused by a dusty magnetospheric stream \citep{Bodman2017,re19,Nagel2020,Nagel2024,Nagel2025}.
Our sample's bias towards high M values is consistent with the relatively high magnetospheric viewing inclinations (\imag), which are above 45\degrees\ for 74\% of the CTTS \citep{Pittman2025}.
Previous work has found that the outer disk inclinations of dipper sources are isotropic \citep{Ansdell2020}, so our finding that the inner magnetospheric inclinations of dippers are non-isotropic provides further evidence that misalignments may be common between the outer disk, inner disk, stellar rotation axis, and/or magnetic axis.

The propeller category has a median M value ($M_{\rm med}$) of 0.25, indicating a high level of ``dippiness'' (see Figure~\ref{fig:QM_violin}, left).
Four out of five propellers are classified as a dipper in at least one TESS sector, and these have $i_{\rm mag}\geq54^\circ$.
These align with the disk warp paradigm expected for systems with \ri$\sim$\rco\ and $i\gtrsim60^\circ$.
The only propeller that is never classified as a dipper (DM~Tau) has a low \imag\ of 22\degrees, which is unlikely to permit occultation by disk material. Instead, DM~Tau tends to be classified as a burster, consistent with its low inclination and high accretion rate \citep{Pittman2025}.
While the disk warp and dusty magnetosphere models typically assume that \ri$\sim$\rco, we find dippers at all values of \omegas, indicating that this assumption may be unnecessary.
In future work, we will measure inner dust disk truncation radii, \rd, to examine whether \ri\ is near \rd\ in these dippers, or if another mechanism is responsible for the occultation events \citep[such as the infall of cool outer-disk material from failed disk winds, as suggested by][]{Takasao2022}.

While periodicity is expected to increase with viewing inclination \citep[e.g.,][]{Robinson2021}, we find no such correlation between Q and \imag\ (see Figure~\ref{fig:QM_violin}, right). This may be because the higher-inclination sources tend to be dippers. While starspots and accretion hotspots can create clear, sinusoidal periodicity, variability by occultation can be more variable over different phases due to changing magnetospheric dust densities or disk warp structures \citep{McGinnis2015,Alencar2010}.
This may be one cause of the high median Q value of the propeller systems ($Q_{\rm med}$=0.76), which also have the highest $M_{\rm med}$ value. 

Interestingly, CTTS in the stable accretion regime tend to be much less periodic than those in the unstable regimes.
As shown in Figure~\ref{fig:QM_violin} (top), the typical Q metrics of the stable regime ($Q_{\rm med}$=0.72) are comparable with those of the 6 CTTS without confident period detections ($Q_{\rm med}$=0.71). In contrast, the unstable chaotic and unstable ordered regimes have $Q_{\rm med}$ of 0.58 and 0.55, respectively.
The dipper/burster occurrence is not significantly different between the stable and unstable regimes, so light curve symmetry does not appear to explain the periodicity differences.
There are also no notable differences in the \imag\ distributions of the accretion stability regimes.
One possible explanation for the lower level of periodicity in the stable regime is emission from an axisymmetric hotspot, which would produce minimal rotational modulation. This could occur if the higher level of stability of the inner disk permits more balanced mass-loading onto the stellar field lines. 
More observations are necessary to confirm light curve morphology differences between the different regimes, though, as this sample is too small to enable statistically robust comparisons.


\subsection{Magnetospheric truncation and ultra-short-period exoplanets} \label{subsec:exoplanets}

Most exoplanets have periods longer than 10~days, with the occurrence rate of shorter-period planets decreasing sharply below this point \citep[e.g.,][]{Youdin2011,Mulders2015}.
However, there is a non-negligible population of exoplanets with orbital periods (\porb) less than 1~day, called ultra-short-period planets (USPs). These have occurrence rates of 0.5--1.1\% around G-, K-, and M-type hosts \citep[][]{Sahu2006,SanchisOjeda2014} and typically have radii below 2~$R_\oplus$, either due to initially small sizes or atmospheric escape due to the strong stellar irradiation \citep{Winn2017,Winn2018}.

In the approximately Earth-sized regime of USPs, planets are expected to form within the disk's snow line, and then migrate inwards due to Lindblad torques driven by the planet-disk interaction \citep{Ward1997}.
If the planet is to survive, there must be a mechanism to halt migration before the planet accretes onto the star.
Surface density gradients in the disk are a leading candidate for stalling planet migration, and there are two primary possibilities: \ri, which marks a strong density transition in the inner gas disk, and the dust sublimation radius, \rd, which marks a strong transition in the dust density and temperature profile \citep{Masset2006, LeeChiang2017, Batygin2023}.

Analytically, \ri\ is expected to be either weakly dependent on \mstar\ \citep[${\rm R}_i\propto{\rm M}_\star^{-1/7}$,][]{hartmann16} or essentially independent of \mstar\ \citep[][]{Batygin2023}. Conversely, \rd\ is expected to be directly correlated with \mstar, as the higher luminosity of higher-mass stars will cause the dust to sublimate at larger radii. 
The semi-major axes of hot Jupiters have been found to be independent of stellar mass, indicating that \ri\ is the trap for these planets \citep{Mendigutia2024}. This is also supported by 3D MHD models of planet migration and stalling \citep[][]{CevallosSoto2025}.
In contrast, the majority of terrestrial planets are located beyond $\sim$0.1~au \citep[][]{Mulders2018}, which is closer to typical values of \rd\ than \ri\ \citep[as \rd\ is usually larger than $\sim$0.06~au, see][]{muzerolle03,Eisner2005,Anthonioz2015,pittman22}. Furthermore, the semi-major axes of the innermost terrestrial planets may be correlated with stellar mass, providing additional evidence for formation or trapping near \rd\ \citep[][]{Mulders2015,Sun2025}.
However, the USP terrestrial planets are typically below $\sim$0.02~au, in the regime of \ri. This requires a different mechanism for formation than for other short-period terrestrial planets.

Until now, short-period planet formation models have assumed magnetospheric truncation radii values by either a) co-locating \ri\ with \rco\ \citep[that is, assuming disk locking; e.g.,][]{LeeChiang2017}, or b) estimating \ri\ analytically using theoretical (and approximate) relationships between the stellar mass, radius, accretion rate, and magnetic field strength \citep[e.g.,][]{Batygin2023}. These assumptions for \ri\ place proto-USPs at radii larger than the observed USP locations and therefore require additional angular momentum loss mechanisms to bring the planets closer to the host stars.
For example, \cite{LeeChiang2017} require tidal dissipation to act for 5~Gyr to explain current USP observations, and \cite{Becker2021} argue that extended enhanced accretion events (such as FU~Ori outbursts) might be required to introduce aerodynamic drag from sub-Keplerian gas motion and bring planets from $>$0.05~au down to 0.01--0.02~au.

\begin{figure}[t]
    \centering
    \includegraphics[width=0.9\linewidth]{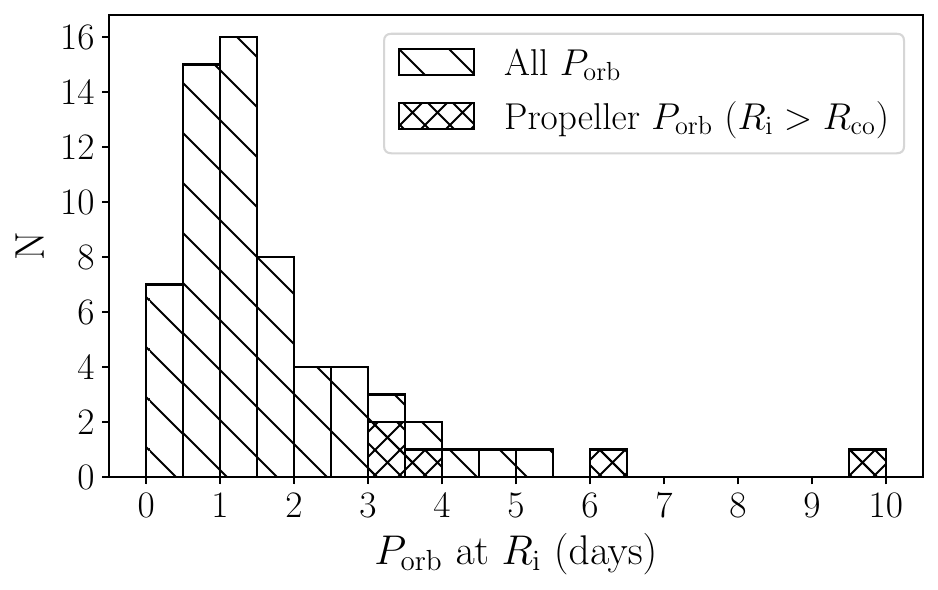}
    \includegraphics[width=0.9\linewidth]{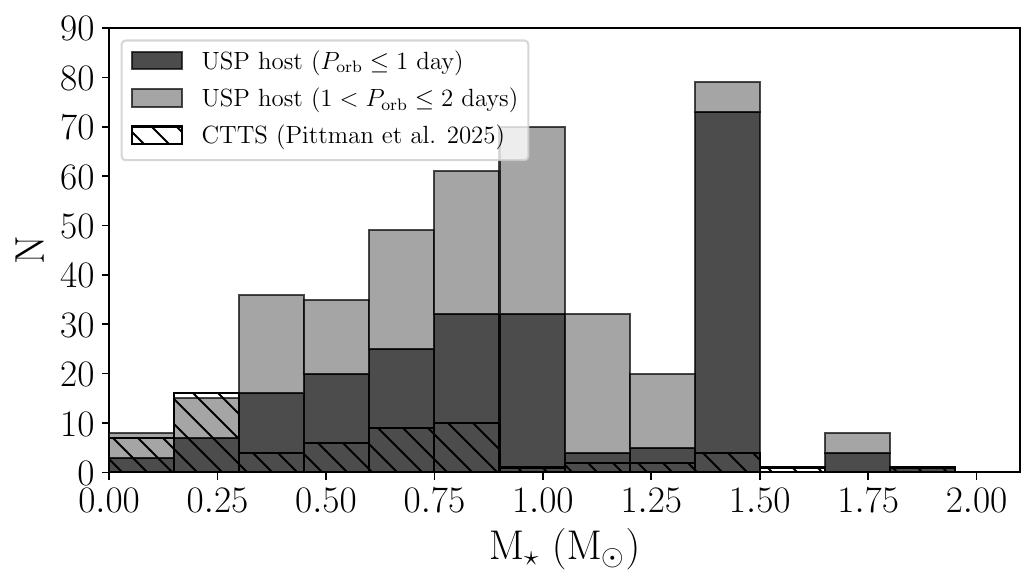}
    \includegraphics[width=0.9\linewidth]{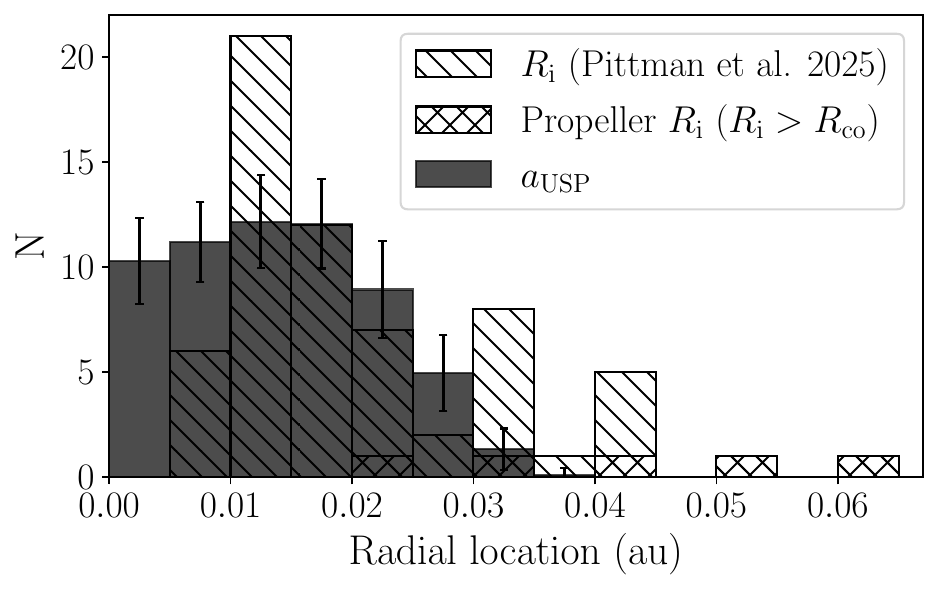}
    \caption{\textit{Top}: Keplerian orbital periods at \ri\ for our sample of CTTS. Values for systems in the propeller regime (where \ri$>$\rco) are indicated.
    \textit{Middle}: Stellar mass distributions of the CTTS (hatched) compared to the hosts of USPs with $P_{\rm orb}\leq1$~day (filled dark gray) and $1<P_{\rm orb}\leq2$~days (filled light gray).
    \textit{Bottom}: Average observed USP semi-major axes ($a_{\rm USP}$, filled) compared to the measured CTTS \ri\ values (hatched). The error bars indicate the standard deviation of $a_{\rm USP}$ in that bin.}
    \label{fig:USPs}
\end{figure}

We have found these \ri\ assumptions to be overly simplistic, as a) \ri\ is notably smaller than \rco\ in most CTTS, and b) analytical calculations of \ri\ rely on the assumed magnetic field strength, and this has been measured for only $\sim$10 CTTS. When calculating \ri\ analytically, it is typical to assume a polar dipole field strength of 1~kG for all CTTS. However, observed \Bpolar\ values vary over time in individual CTTS and range from $\sim$0.08--1.72~kG, and even up to 3~kG depending on the tracer used \citep[][and references therein]{Donati2010,Grankin2021,Johnstone2014}.

Similarly, using Equation~\ref{Eq:B} with $B_{\rm \star,eq} = 0.5B_{\rm \star,polar}$ gives \Bpolar\ values that range from 0.06--6.57~kG for the \cite{Pittman2025} CTTS sample, with a median of 0.7~kG and a mean of 1.2~kG. Thus, assuming a fiducial \Bpolar\ value for all CTTS cannot produce \ri\ values that are in agreement with our measured values due to intrinsic magnetic field strength and efficiency differences between different CTTS (see the discussion in \citealt{thanathibodee23}).\footnote{We note that four CTTS in our sample (AA~Tau, DN~Tau, CV~Cha, and Sz~75) have measured dipole field strengths in \cite{Donati2010,Donati2012,Donati2013} and 
\cite{Johnstone2014}. However, magnetic obliquity and temporal variability make it non-trivial to compare \Bpolar\ measured from ZDI surface field maps with our \Beq\ measured at different epochs under an axisymmetric dipole approximation.}

Using the accretion flow model, we find smaller \ri\ values than have been previously assumed, with a median value of 0.016~au. The corresponding Keplerian orbital periods at \ri\ in our sample range from 0.2--9.8~days, with a median of 1.3~days (see Figure~\ref{fig:USPs}, top).
To compare \ri\ with the observed USP semi-major axes ($a_{\rm USP}$), we adopt a sample of confirmed exoplanets with $a_{\rm USP}$ measurements around host stars with $M_\star<2$~\msun\ from the Extrasolar Planets Encyclopaedia.\footnote{\url{https://exoplanet.eu/}}$^{,}$\footnote{When the stellar mass is unavailable, we estimate \mstar\ using \porb\ and $a_{\rm USP}$ using Kepler's law. This is reasonable because short-period planets typically have low eccentricities \citep{Kipping2013}.} 
It is important for the USP hosts to have a comparable \mstar\ distribution as our CTTS sample, as $a_{\rm USP}$ is dependent on \mstar.
If we only include USPs with $P_{\rm orb}<1$~day, the host \mstar\ distribution lacks a sufficient number of stars with $M_\star<0.3$~\msun\ (see Figure~\ref{fig:USPs}, middle).
The historical one-day cutoff that defines USPs is arbitrary \citep[though there may be physically-motivated evidence for both a one- and two-day cutoff, see][]{GoyalWang2025}, so we choose a cutoff of two days to permit an \mstar-balanced comparison.

We perform bootstrapping to select subsets of the observed USPs with distributions of host \mstar\ that are consistent with our CTTS \mstar\ distribution.
We calculate $a_{\rm USP}$ histograms for 5000 bootstrapped samples, each with the same number of USPs as our CTTS sample. We then calculate the mean and standard deviation of each bin. This is shown in Figure~\ref{fig:USPs} (bottom), along with our measured CTTS \ri\ values. The median observed USP location is 0.016~au, equivalent to our measured \ri\ median of 0.016~au. The shapes of the distributions differ, but the rarity of USPs means that only a subset of CTTS need to have favorably-located magnetospheric truncation radii to produce the observed USP occurrence rates.

Our results demonstrate that observed USP semi-major axes are aligned with measured magnetic truncation radii. However, the majority of these overlapping values correspond to systems that are in the unstable regime (in which \ri$\ll$\rco), and therefore the local Keplerian velocity is significantly faster than the stellar rotation.
\citet[][Equation 1]{Ogilvie2014} gives the critical angular momentum required to support tidal equilibrium in a given system. We compared the observed and critical angular momentum values for each CTTS and found that, while planets with masses up to 0.2~$M_{\rm Jup}$ have tidal equilibrium solutions for our entire sample, the solutions are in the too-compact unstable branch that will eventually decay given any perturbation. 
However, \cite{Hamer2020} find observationally that tidal dissipation efficiency in USP systems is low, which allows USPs to avoid tidal decay into the host stars over their main-sequence lifetimes.
Given this context, we suggest that while Earth-mass planets orbiting at these measured \ri\ values would be theoretically unstable to tidal dissipation, the dissipation efficiency may be so low that it does not cause tidal decay over the stellar main sequence lifetime.

\cite{Schmidt2024} and \cite{Tu2025} used stellar velocity dispersions to find that USPs tend to be around older host stars, indicating that they often form later in the system evolution. However, there may be systematic differences between younger and older USP populations that imply multiple pathways for USP formation \citep{Tu2025}.
The range of \ri\ values we measure is consistent with a subset of proto-USPs arriving at their final locations by the time the disk dissipates, while others stall at \ri\ values that correspond to longer periods and must slowly evolve to the ultra-short periods over a few Gyr timescales.
Future work should examine the implications of USP formation through direct trapping at \ri\ without requiring additional orbital decay processes (such as planet-planet scattering and tidal decay). It should also determine the consequences of short-timescale \ri\ variability (discussed in Section~\ref{subsec:RiVar_short}) on the stability of USP orbits.

\section{Summary} \label{sec:summary}

In this work, we have examined the star-disk connection in the well-characterized HST ULLYSES sample.
In \cite{Pittman2025}, we applied accretion flow models to velocity-resolved \halpha\ profiles to measure magnetospheric truncation radii (\ri) for 66 CTTS.
Here, using TESS light curves, we measured stellar rotation periods and corotation radii (\rco) for a subset of 47 CTTS.
We compared \ri\ and \rco\ to study the spin states of the sample.
Our results can be summarized as follows:

\begin{enumerate}
    \item We classified each CTTS into the expected accretion stability regimes based on the fastness parameter, \omegas=$(R_{\rm i}/R_{\rm co})^{3/2}$. We found that individual CTTS cross regime boundaries as part of their dynamic accretion processes.
    \item Using the star-disk interaction torque relation from \cite{Zhu2025}, we find that most systems are in the spin-up regime, rather than being in spin equilibrium. This is significant because both theoretical work and observational population studies of CTTS often assume rotational equilibrium. Future work should explore the cause of the observed spread in CTTS spin states.
    \item Blueshifted absorption features indicative of conical winds launched at \ri\ are more common in CTTS with higher \omegas, consistent with expectations for this mechanism of angular momentum regulation.
    \item CS~Cha shows evidence of an episodic accretion event in which a magnetospheric expansion was accompanied by a decrease in \mdot\ and the appearance of a wind. 
    \item Our observations are consistent with the ability of accretion-powered stellar winds to carry away 20--30\% of the accretion torque.
    \item We quantified the periodicity and symmetry of TESS light curves and connected these with the accretion stability regimes and system characteristics. Our findings generally agree with previous work. Dippers tend to be viewed at inclinations larger than $\sim$50\degrees, whereas bursters tend to be viewed at lower inclinations and have accretion rates above $6\times10^{-9}$~\msunyr. Systems in the propeller regime tend to be more stochastic than periodic, and they are often dippers, consistent with disk warps being driven by \ri$\sim$\rco\ in those systems. 
    However, we find dippers at all values of \omegas, suggesting that disk warps and/or dusty magnetospheres may appear in non-propeller systems as well.
    \item We found that measured \ri\ values show good correspondence with the observed semi-major axes of ultra-short period exoplanets (USPs). If USPs are stable against tidal dissipation, then migration halting at \ri\ is a promising mechanism for USP formation. 
\end{enumerate}

These results show that common simplifying assumptions made with respect to stellar magnetospheres and angular momentum evolution | specifically, that magnetospheres are uniform in size, stable, and in spin equilibrium with the inner disk | are inconsistent with observations.
The large range of observed \omegas\ values implies a complex star-disk interaction, and this motivates continued investigation of the implications of time-variable, non-axisymmetric, three-dimensional stellar magnetospheres on star and planet formation and co-evolution.

\begin{acknowledgments}

We wish to recognize the work of Will Fischer, who passed away in 2024. His dedication at STScI ensured the successful implementation of ULLYSES, and his kindness made him a friend to many. 

We sincerely thank the anonymous reviewer for their insightful feedback that improved the manuscript.
We thank Silvia Alencar, Antonio Armeni, Jerome Bouvier, Eric Gaidos, Lukas Gehrig, Uma Gorti, Konstantin Grankin, Lee Hartmann, Shinsuke Takasao, and Johanna Teske for valuable discussion about this work.
We thank Javier Serna for the continuous maintenance of \texttt{TESSExtractor} and for providing assistance with interpreting its outputs.

This work was supported by HST AR-16129 and benefited from discussions with the ODYSSEUS team \citep[\url{https://sites.bu.edu/odysseus/},][]{espaillat22}. C.P. acknowledges funding from the NSF Graduate Research Fellowship Program under grant No. DGE-1840990.
This work was also supported by the NKFIH NKKP grant ADVANCED 149943 and the NKFIH excellence grant TKP2021-NKTA-64. Project no.149943 has been implemented with the support provided by the Ministry of Culture and Innovation of Hungary from the National Research, Development and Innovation Fund, financed under the NKKP ADVANCED funding scheme.
Funding for the TESS mission is provided by the NASA Science Mission Directorate. 
TIis research has made use of data obtained from or tools provided by the portal \url{exoplanet.eu} of The Extrasolar Planets Encyclopaedia.

\end{acknowledgments}

\facilities{TESS, VLT(ESPRESSO)}

\software{astropy \citep{2013A&A...558A..33A,2018AJ....156..123A,Astropy2022}, TESSExtractor \citep{brasseur19,serna21}, scikit-learn \citep{scikit-learn}}

\bibliography{bib}{}
\bibliographystyle{aasjournal}

\appendix
\section{Magnetospheric configurations from the best-fit accretion flow models} \label{Appsec:app_flowresults}

Here we show the magnetospheric configurations of the full sample from \paperI. Systems with \rco\ measurements are ordered by decreasing \omegas, and then those without \rco\ measurements are ordered by region (youngest to oldest). 
We show the following parameters to scale: \rstar\ (R$_{\star,f}$ in Table~1 in \citealt{Pittman2025}), \teff, \tshock\ (from the initial iteration of the accretion shock modeling in \citealt{Pittman2025}), \tmax, \ri, \rw, and \rco. The figures also include a rough estimate of the dipole magnetic field strength at the stellar equator, \Beq, calculated according to Equation~\ref{Eq:B}. Note that this is half the value at the stellar pole (\Bpolar) for a pure dipole, assuming that the magnetic axis is parallel to the stellar rotation axis. The inclination from the magnetic axis relative to the observer, \imag, is indicated by the black arrow.
As discussed in Section~\ref{subsec:ProtReliability}, the geometry of the flow model assumes that \imag\ is parallel to the stellar rotation axis and perpendicular to the plane of the inner disk, represented by the shaded gray region.
The color bar uses the color scheme from \cite{harreheller21} to convert \teff\ to visual color to demonstrate the contrast between the stellar, accretion hotspot, and maximum accretion flow temperatures for each system. The horizontal axis extends to 0.1~au, which is the most common location of the short-period ``Kepler planets'' \citep{Mulders2018}.

\begin{figure*}
    \centering
    \includegraphics[width=0.93\textwidth]{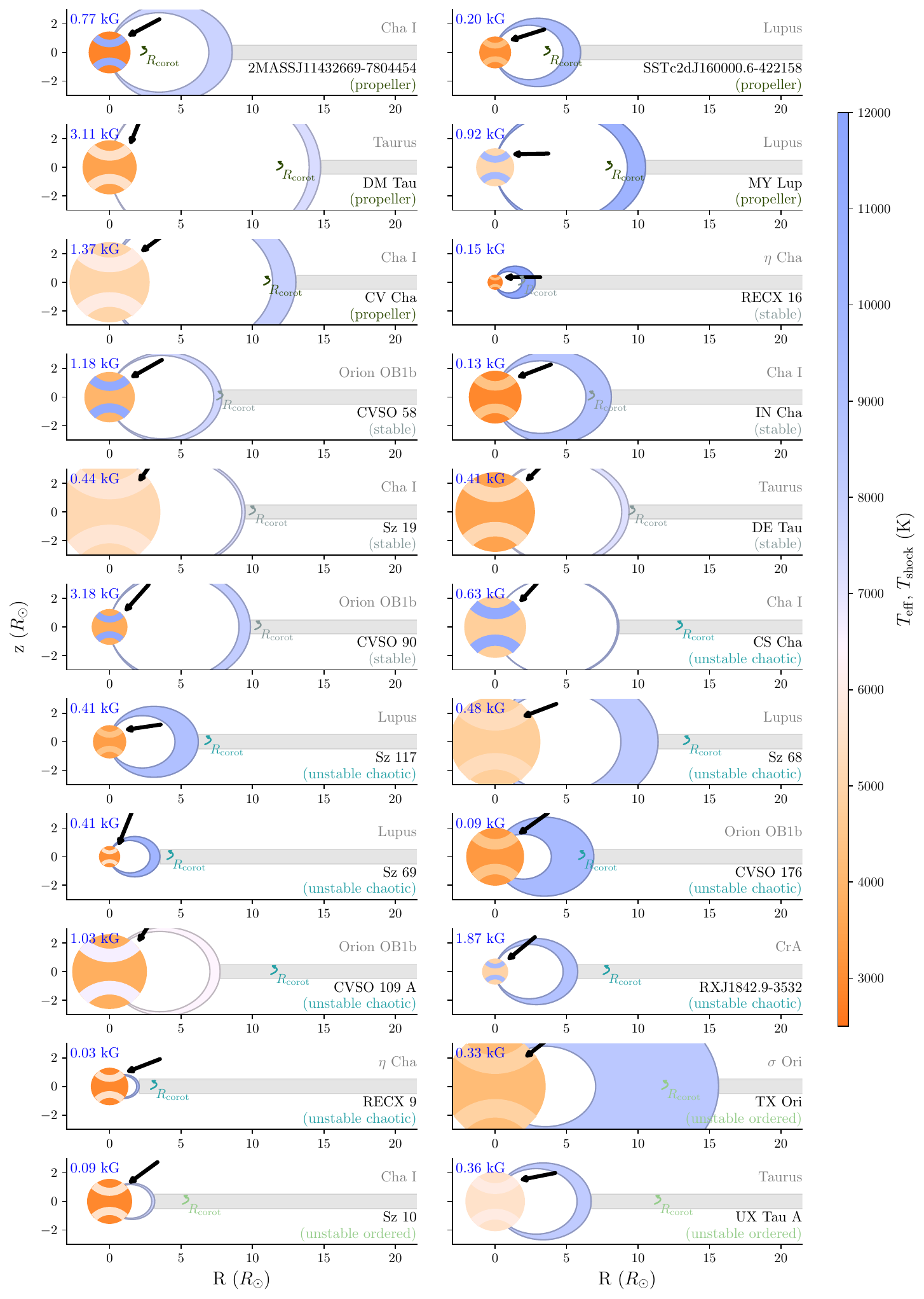}
    \caption{System configurations. See text for details.}
    \label{fig:RadiusFigure_twocol0}
\end{figure*}

\begin{figure*}
    \ContinuedFloat
    \centering
    \includegraphics[width=0.93\textwidth]{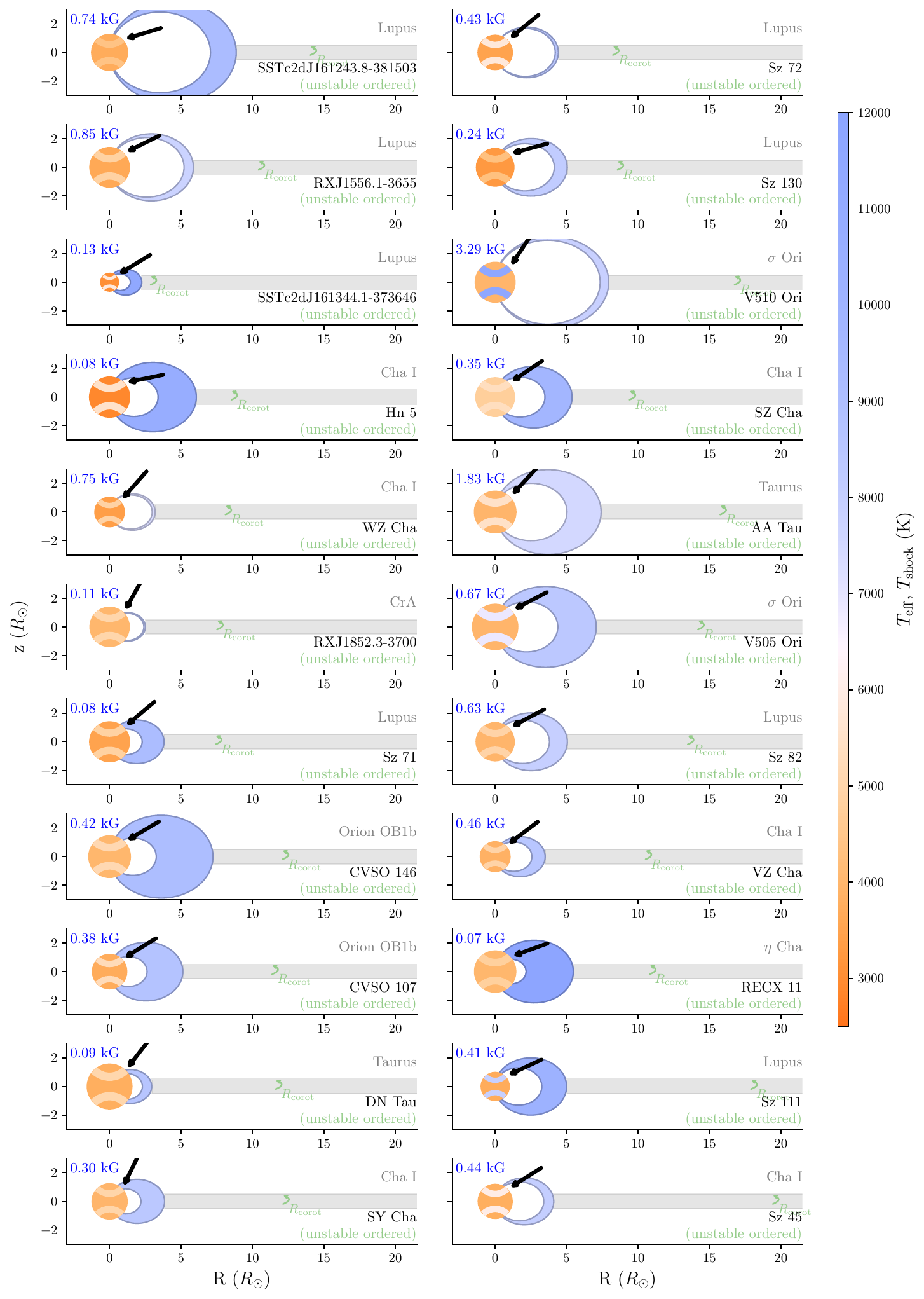}
    \caption{System configurations continued.}
    \label{fig:RadiusFigure_twocol1}
\end{figure*}

\begin{figure*}
    \ContinuedFloat
    \centering
    \includegraphics[width=0.93\textwidth]{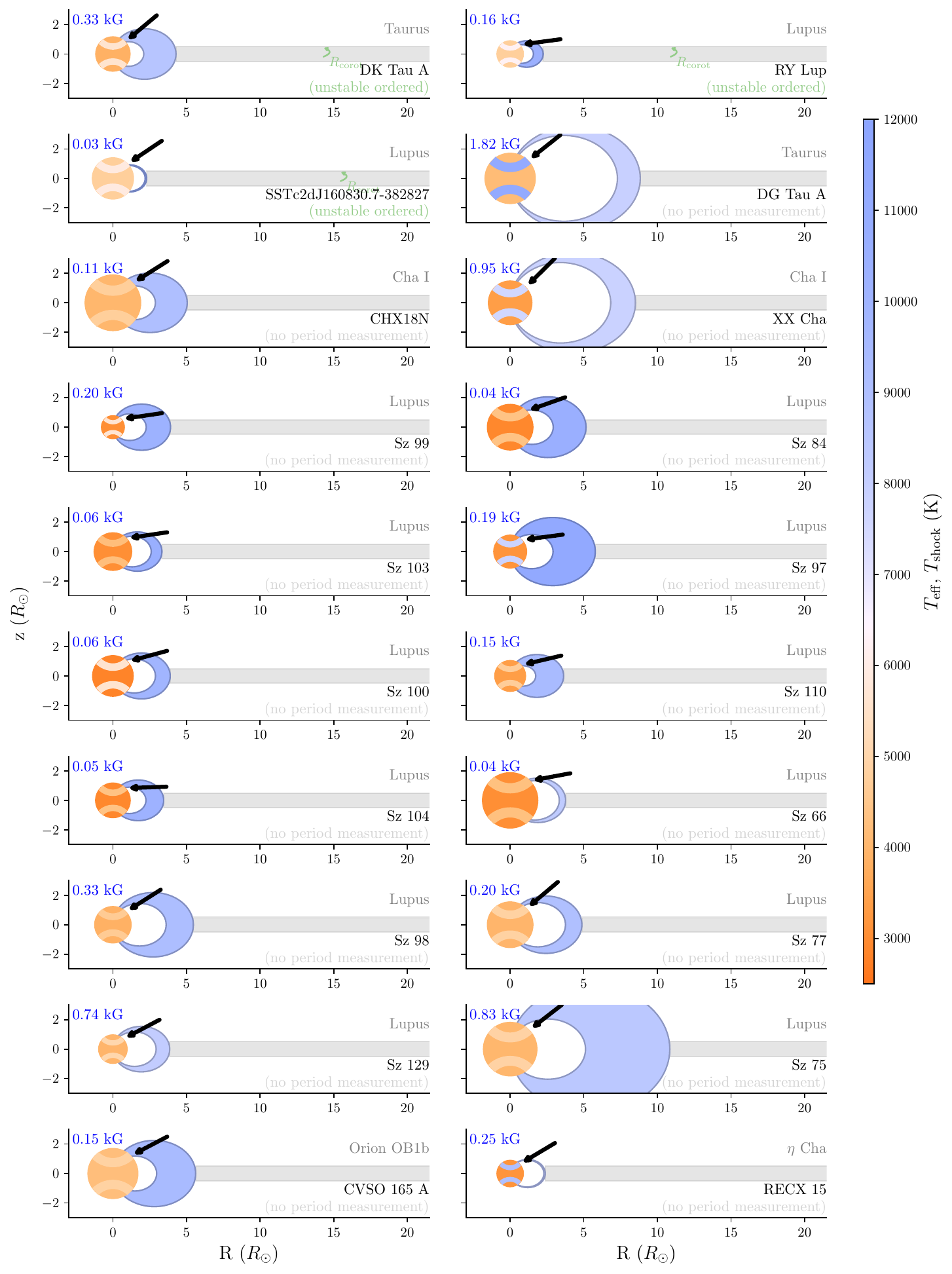}
    \caption{System configurations continued. Rows DG~Tau~A and below have no \prot\ measurements (and therefore no \rco\ or \omegas).}
    \label{fig:RadiusFigure_twocol2}
\end{figure*}

\section{Period validation} \label{Appsec:per_val}

\begin{figure*}
    \centering
    \includegraphics[width=0.6\textwidth]{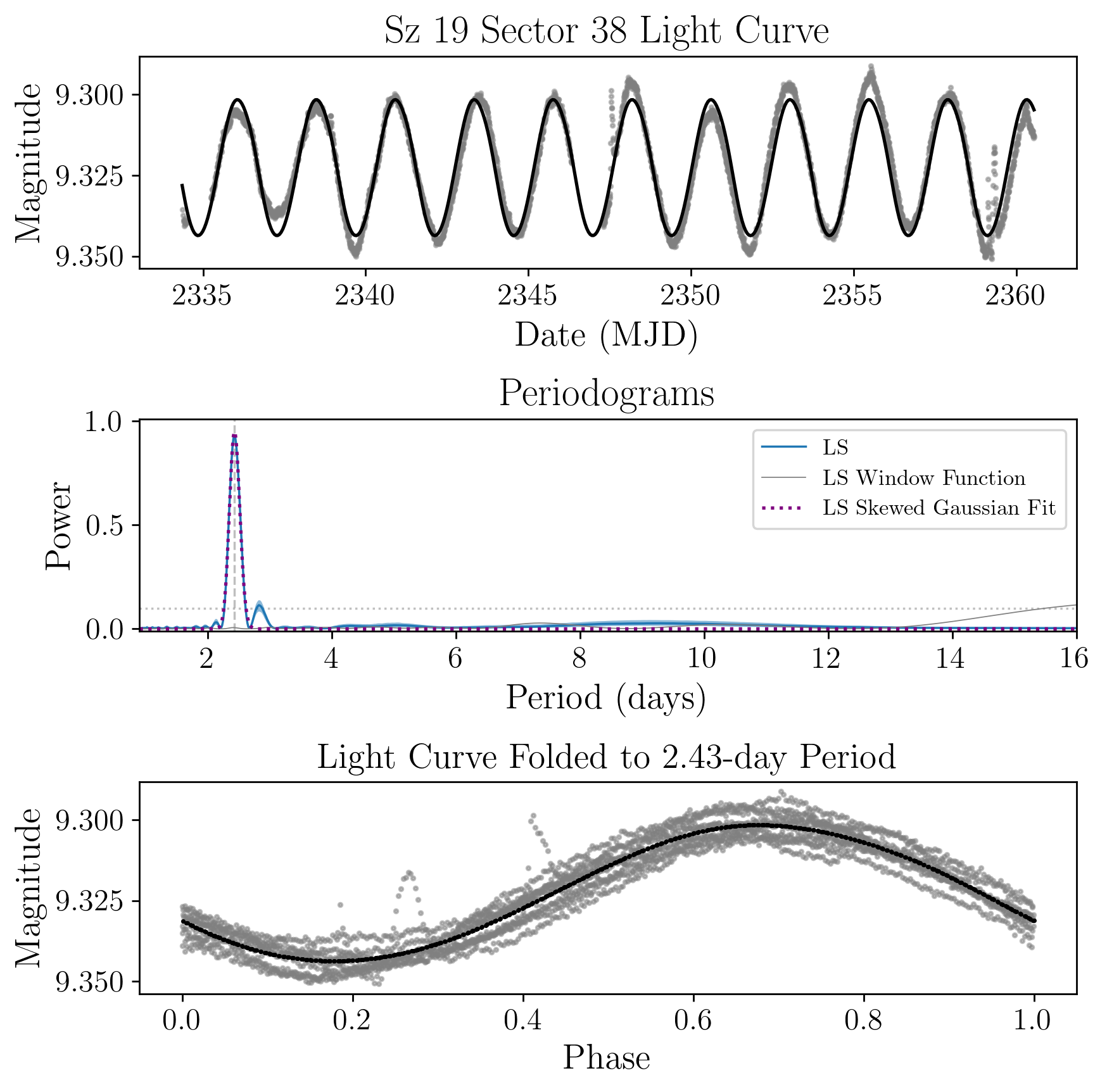}
    \includegraphics[width=0.65\textwidth]{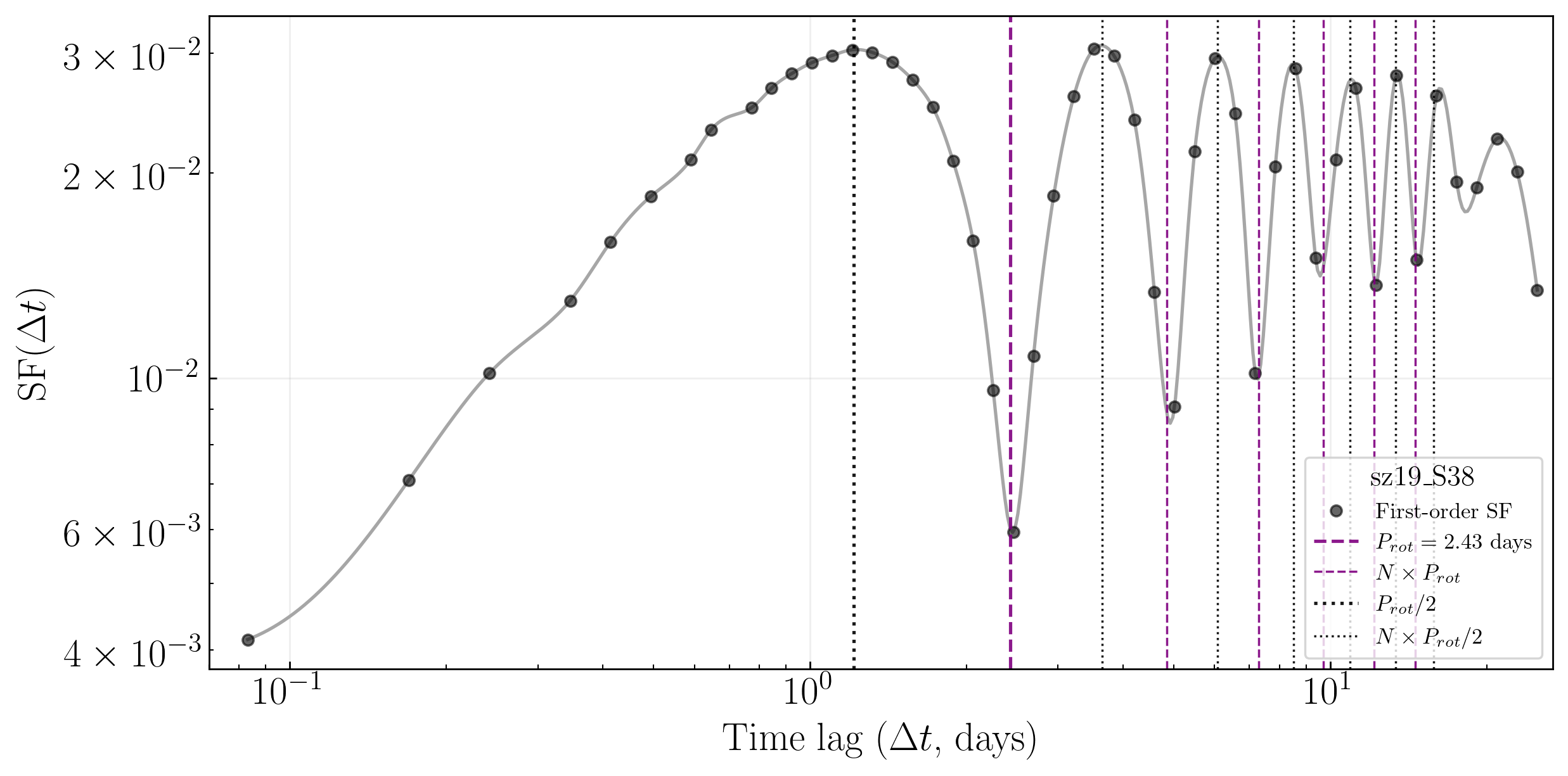}
    \caption{Example period measurement method for Sz~19 Sector 38. The first panel shows the TESS light curve (gray) and its best-fit Gaussian process with a periodic kernel (black). The second panel shows the Lomb-Scargle periodogram (blue), the window function periodogram (gray), and the skew-normal function fit to the peak to measure the period uncertainty. The third panel shows the phase-folded light curve. The final panel shows the structure function (SF) with multiples of \prot\ and \prot/2 indicated. The SF is low at integer multiples of \prot\ and high at half-integer multiples of \prot.}
    \label{fig:LS_SF}
\end{figure*}

We measure and validate periods from TESS light curves as described in Section~\ref{sec:analysis}. Here we show an example in Figure~\ref{fig:LS_SF}.

\section{Additional QM results} \label{Appsec:additionalQM}

\begin{figure*}
    \centering
    \includegraphics[width=0.9\linewidth]{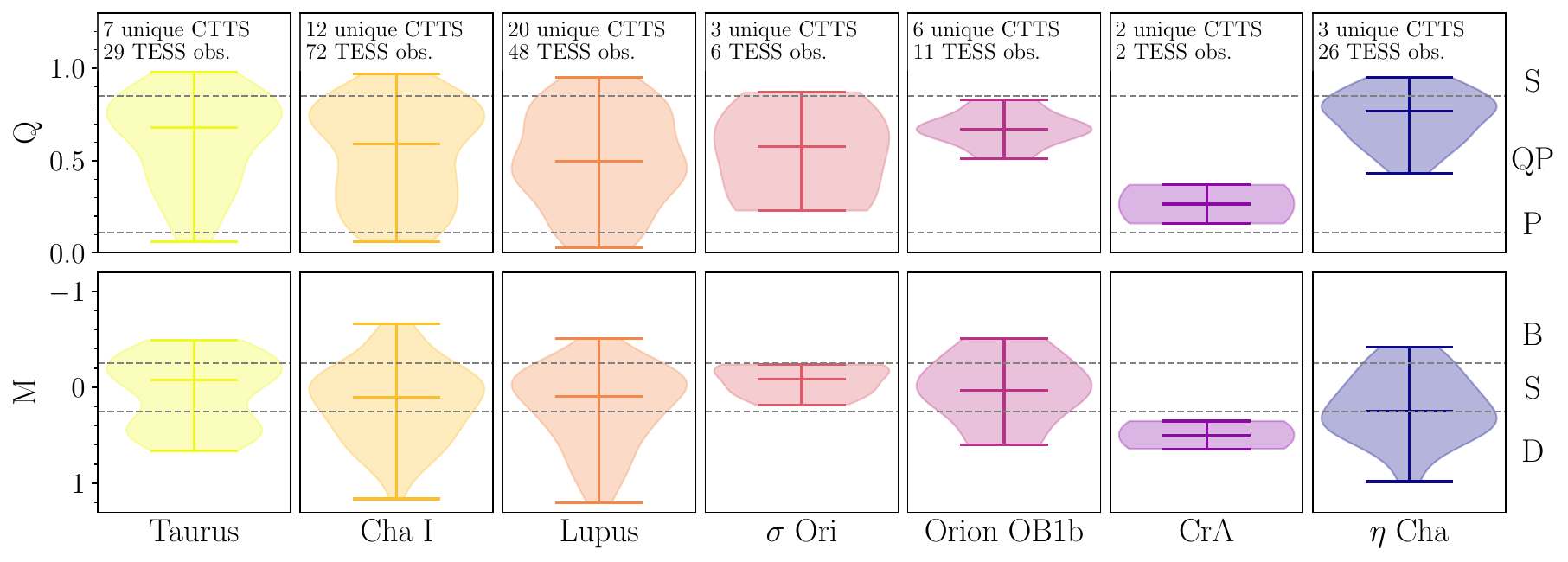}
    \includegraphics[width=0.5\linewidth]{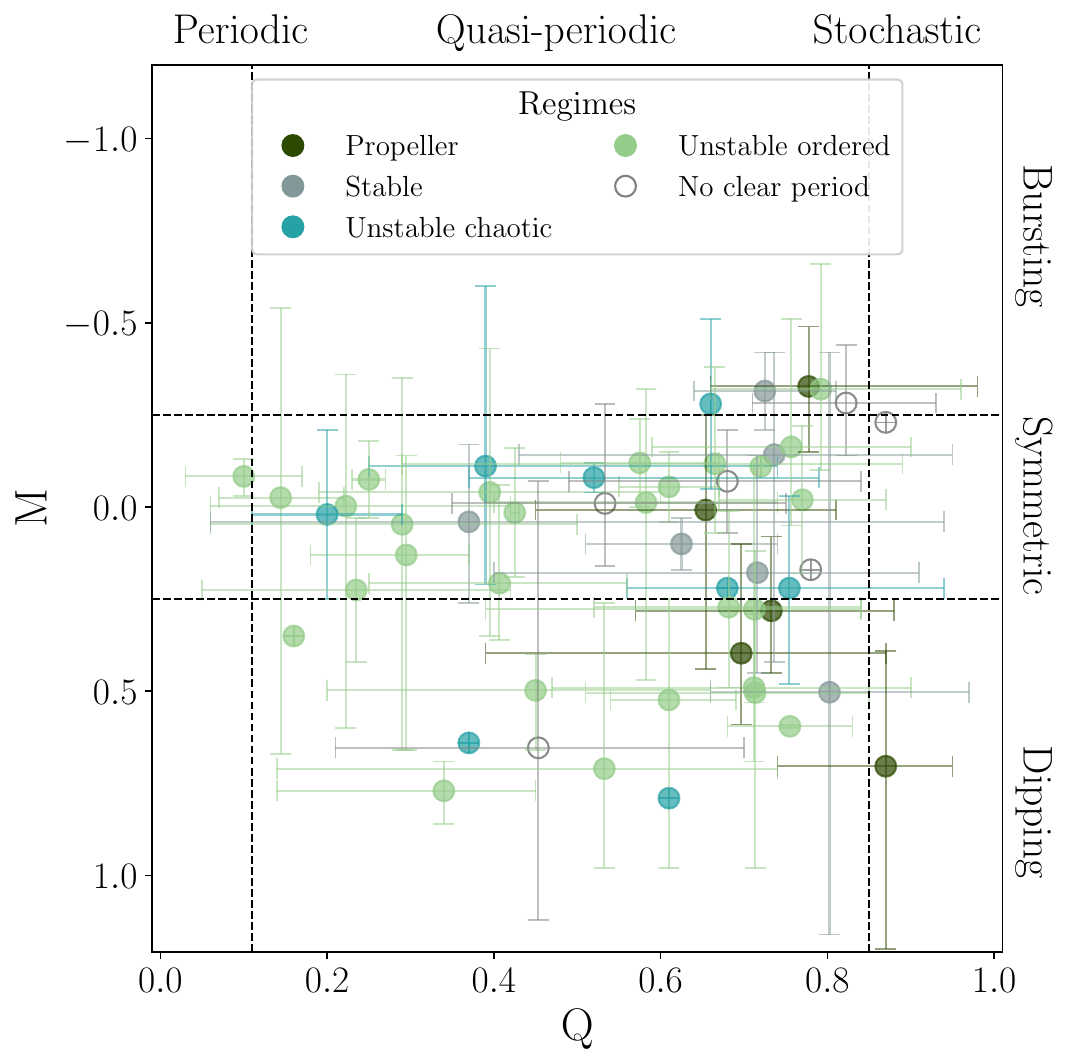}
    \caption{
    Additional QM variability results. \textit{Top:} Results for all targets with good TESS data (whether or not a strong period was detected), grouped by host region in order of median region age \citep{Michel2021}. The minimum, median, and maximum of each distribution is indicated by the horizontal solid lines. 
    Gray dashed lines indicate the divisions between categories as described in Section~\ref{sec:analysis}, with labels on the far right indicating the Q categories (S/Stochastic, QP/Quasi-Periodic, P/Periodic) and M categories (B/Burster, S/Symmetric, D/Dipper).
    \textit{Bottom}: QM scatterplot colored according to accretion stability regime, where the circle is marked at the average value of Q and M. The bars indicate the full range of Q and M for each target across all TESS sectors.
    }
    \label{fig:additionalQM}
\end{figure*}

\cite{Cody18} found a higher dipper fraction in Upper Scorpius compared to younger star forming regions, suggesting an age dependence. Figure~\ref{fig:additionalQM} (top) shows the QM results separated by host region in order from youngest (Taurus) to oldest ($\eta$~Cha). A statistically robust comparison is not possible because the older regions have significantly fewer targets, but the limited data available are consistent with some age dependence.
Figure~\ref{fig:additionalQM} (bottom) shows our results on the QM plane for easier comparison with previous work that has applied the \cite{Cody14} methodology. The Q metric boundaries are the same as those of \cite{Cody18}, as discussed in Section~\ref{subsec:VariabilityMetrics}.

\section{Long-term TESS light curve morphology changes} \label{Appsec:MorphologyChanges}

\begin{figure*}
    \centering
    \includegraphics[trim=7pt 7pt 7pt 6pt, clip, width=0.26\textwidth]{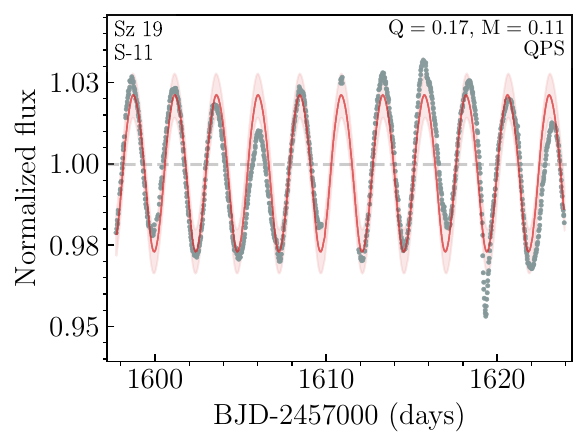}
    \includegraphics[trim=22pt 7pt 7pt 6pt, clip, width=0.24\textwidth]{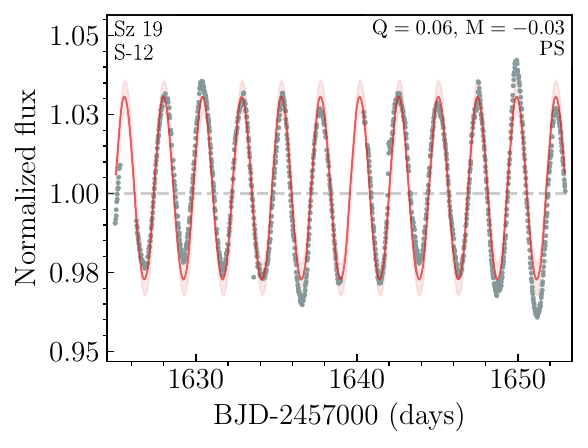}
    \includegraphics[trim=22pt 7pt 7pt 6pt, clip, width=0.24\textwidth]{figures/sz19_S38_lc.pdf}
    \includegraphics[trim=22pt 7pt 7pt 6pt, clip, width=0.24\textwidth]{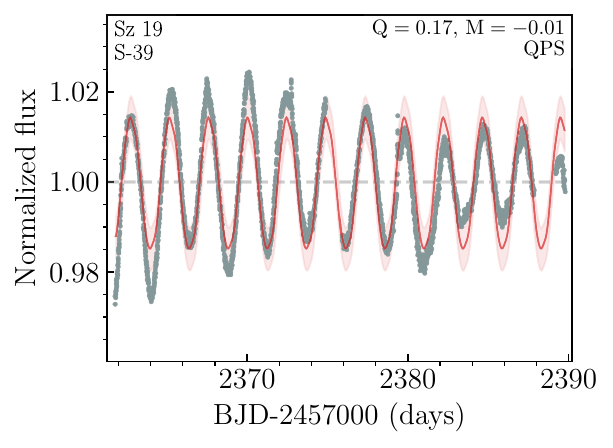} \\
    \includegraphics[trim=7pt 7pt 7pt 6pt, clip, width=0.26\textwidth]{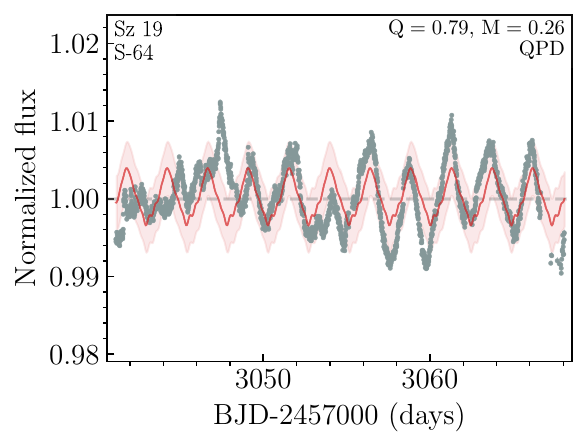}
    \includegraphics[trim=22pt 7pt 7pt 6pt, clip, width=0.24\textwidth]{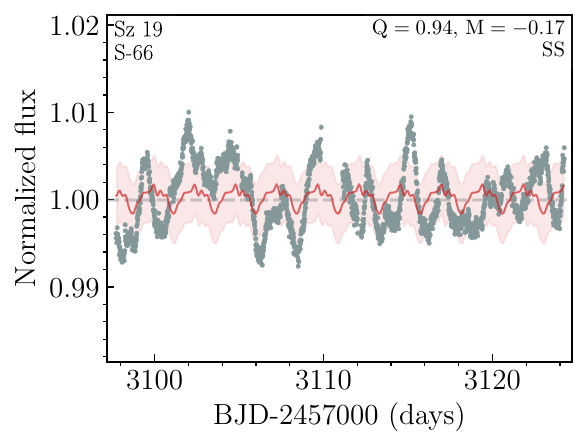}
    \includegraphics[trim=22pt 7pt 7pt 6pt, clip, width=0.24\textwidth]{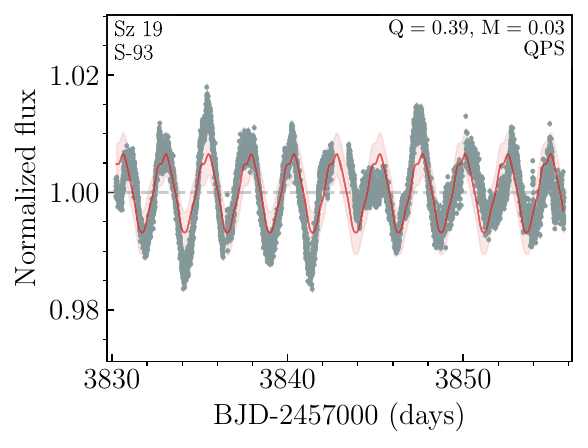}
    \caption{TESS light curves of Sz~19, which switches from highly periodic in 2019/2021 (top) to quasi-periodic and stochastic in 2023 and 2025 (bottom). The associated QM metrics are in the upper right, and the TESS sector is in the upper left. A Gaussian process with a periodic kernel fixed to 2.43~days is shown in red, and its standard deviation is indicated by the shaded red region.} 
    \label{fig:Sz19_TESS}
\end{figure*}

\begin{figure*}
    \centering
    \includegraphics[trim=7pt 7pt 7pt 6pt, clip, width=0.26\textwidth]{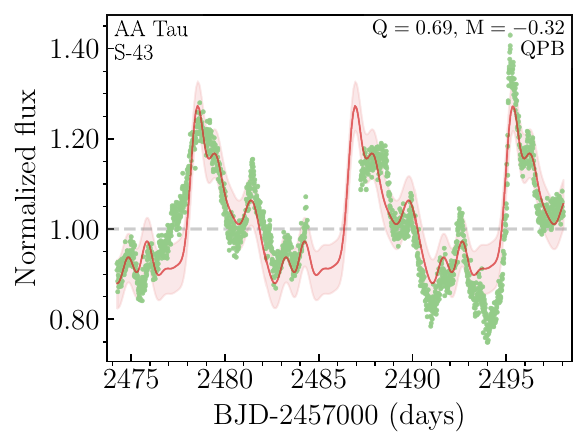}
    \includegraphics[trim=22pt 7pt 7pt 6pt, clip, width=0.24\textwidth]{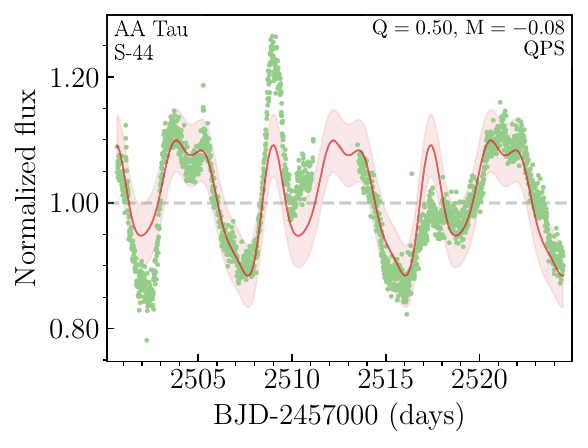}
    \includegraphics[trim=22pt 7pt 7pt 6pt, clip, width=0.24\textwidth]{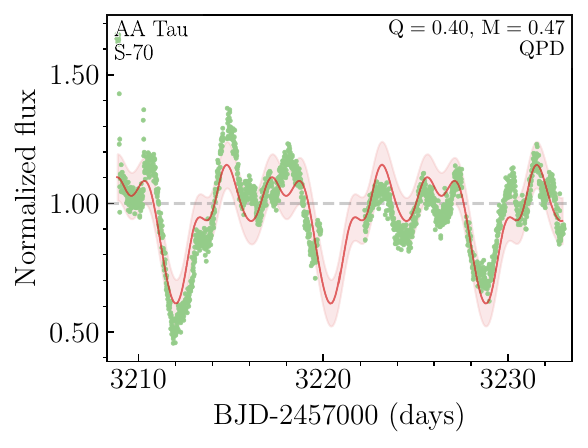}
    \includegraphics[trim=22pt 7pt 7pt 6pt, clip, width=0.24\textwidth]{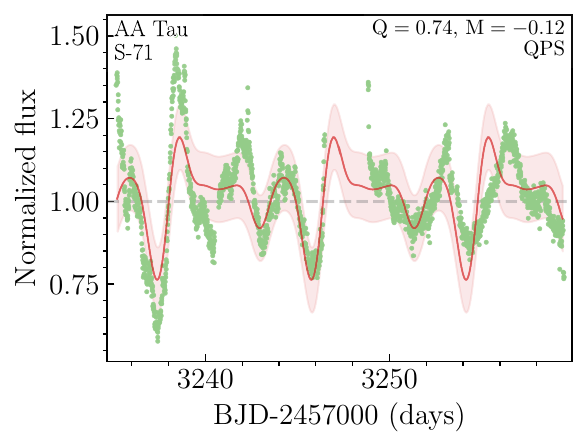}
    \caption{TESS light curves of AA~Tau in 2021 and 2023, which span the categories of dipper, symmetric, and burster. The associated QM metrics are in the upper right, and the TESS sector is in the upper left. A Gaussian process with a periodic kernel fixed to 8.38~days is shown in red, and its standard deviation is indicated by the shaded red region.}
    \label{fig:AATau_TESS}
\end{figure*}

Figure~\ref{fig:Sz19_TESS} shows the TESS observations of Sz~19, with the highly-periodic sectors on top and the more stochastic sectors on the bottom. The level of periodicity (that is, the Q value) is correlated with the peak-to-peak magnitude of variability, suggesting that the morphology differences may result from starspot coverage changes. The 2019 and 2021 observations may have traced the system while it had one large starspot, whereas 2023 and 2025 observations may be tracing multiple smaller spots. Such changes may indicate intrinsic magnetic field changes, but dedicated observations would be required to confirm this hypothesis.
Figure~\ref{fig:AATau_TESS} shows the TESS observations of AA~Tau in 2021 and 2023, which show both bursts and dips.

\end{document}